\documentclass[lettersize,journal]{IEEEtran}
\usepackage{amsmath,amsfonts}
\usepackage{algorithmic}
\usepackage{algorithm}
\usepackage{array}
\usepackage[caption=false,font=normalsize,labelfont=rm,textfont=rm]{subfig}
\usepackage{textcomp}
\usepackage{stfloats}
\usepackage{url}
\usepackage{verbatim}
\usepackage{graphicx}
\usepackage{cite}
\usepackage{epstopdf}
\usepackage[
colorlinks=true,
linkcolor=blue, 
citecolor=blue, 
urlcolor=blue  
]{hyperref}
\begin{document}

\title{Joint Gaussian Beam Pattern and Its Optimization for Positioning-Assisted Systems}

\author{Yuanbo Liu,~\IEEEmembership{Student Member,~IEEE}, Bingcheng Zhu,~\IEEEmembership{Senior Member,~IEEE}, Shuojin Huang,~\IEEEmembership{Student Member,~IEEE}, Han Zhang, Zaichen Zhang,~\IEEEmembership{Senior Member,~IEEE}

\thanks{Y. Liu, B. Zhu, S. Huang, H. Zhang and Z. Zhang are with the National Mobile Communications Research Laboratory, Frontiers Science Center for Mobile Information Communication and Security, Southeast University, Nanjing 210096, China. B. Zhu, H. Zhang and Z. Zhang are also with the Purple Mountain Laboratories, Nanjing 211111, China. }
\thanks{B. Zhu is the corresponding author (e-mail: zbc@seu.edu.cn).}}



\maketitle

\begin{abstract}
Beamforming is a fundamental technology that not only enhances communication efficiency but also lays the foundation for massive multiple-input multiple-output~(MIMO) systems. However, its reliance on accurate channel state information (CSI) estimation introduces significant training overhead and feedback costs, especially in large-scale antenna systems. In this paper, we investigate positioning-assisted beamforming as a competitive alternative to the CSI-based methods, which circumvents the complicated CSI estimation. In particular, we analyze the outage probability of positioning-assisted systems with joint Gaussian beams and derive its closed-form expressions for both two-dimensional~(2D) and three-dimensional~(3D) scenarios. Based on these results, we also derive closed-form expressions for the optimal joint Gaussian beam pattern. The optimal solution is independent of the positioning error distribution in 2D scenarios but depends on it in 3D cases. Subsequently, the asymptotic performance of the approximation error is analyzed. Numerical results verify the derived outage probability expressions, and show the effectiveness of the beam pattern optimization.
\end{abstract}

\section{Introduction}
\IEEEPARstart{A}{S} a fundamental technology in massive multiple-input multiple-output~(MIMO) systems, beamforming provides substantial gains in signal strength, spatial multiplexing capability, and spectral efficiency. Simultaneously, it enhances target recognition accuracy for sensing applications, making it a promising implementation strategy for integrated sensing and communication~(ISAC) systems~\cite{liu2022cramer,zhang2024unsourced,vahdani2025capacity}. However, beamforming in massive MIMO systems is fundamentally constrained by hardware limitations in channel acquisition. Specifically, the scarcity of radio-frequency~(RF) chains degrades channel state information~(CSI) estimation accuracy, while the scaling of antenna arrays induces exponentially growing pilot overhead, jointly straining training resources~\cite{gao2016energy,jafri2025distributed}. Downlink channel estimation is further exacerbated by prohibitively high CSI feedback costs. Moreover, hybrid beamforming architectures require solving computationally intractable non-convex optimization problems, where the computational complexity exhibits superlinear growth with respect to the system dimension~\cite{cheng2023double,li2023integrated,liu2025doa}.

To address these limitations, positioning information for beamforming design was validated through comprehensive system-level analysis based on a two-dimensional~(2D) uniform rectangular array in the fifth generation~(5G) networks~\cite{talvitie2019positioning}. In~\cite{chou2021fast,li2021bidirectional,wang2021joint}, positioning information is exploited to construct a reduced beam search space and improve beam alignment results, which effectively reduces the training overhead. Nevertheless, these approaches still rely on beam training procedures and thus cannot eliminate the associated training cost. In contrast, the approach in~\cite{maiberger2010location} utilizes receiver positioning information to construct transmit beamforming vectors and avoids the need for complex CSI computation and feedback mechanisms, which provides a new paradigm for beamforming systems design. The capacity of a beamformed radio link was derived in~\cite{talvitie2020beamformed} based on the probability density function~(PDF) of the receiver position, while the resulting expressions involve integrals and a detailed analytical characterization remains unexplored.
  
Building upon these analytical frameworks, some studies have explored positioning-assisted beamforming in a variety of practical deployment scenarios. A statistical model for positioning errors was proposed based on vehicular ISAC channel measurements, but it failed to establish the connection between positioning error and communication performance~\cite{zhang2025channel}. The impact of positioning errors on beamforming performance in vehicle to vehicle~(V2V) networks was analyzed in~\cite{kanthasamy2018assessment}, while explicit analytical relationship between positioning errors and performance metrics are not provided. Furthermore, the work in~\cite{chen2017directivity} examined the tradeoff between beam directivity and positioning error tolerance in V2V networks and established a beam-phase lookup table to reduce beam training latency. However, it is constrained to 2D mobility and dichotomous beam coverage scenarios. Beamforming vectors for 2D millimeter-wave unmanned-aerial-vehicle~(UAV) systems were optimized using a model-based online framework, without relying on offline training data or prior information~\cite{yu2022joint}. However, the obtained solutions are near-optimal and depend on numerical iteration. 
\IEEEpubidadjcol
After that, a position-based hybrid beamforming scheme in vehicle-to-infrastructure (V2I) networks was introduced~\cite{song2024position}, and closed-form solutions for the antennas' optimal physical tilt angles was derived. Nevertheless, its analytical framework also remains limited to 2D road geometries and design of the optimal beams still rely on numerical optimization. To address more general three-dimensional~(3D) scenarios, beam directions were computed directly from user position information in 3D scenarios with limited beam patterns~\cite{lu2020positioning}, and a thorough study of optimal beamforming was not provided. In multi-UAV scenarios, optimal beamforming solutions were studied in~\cite{zhu2022multi}, while the beam pattern is restricted by constant-modulus constraints, and the determination of the optimal beamwidth relies on numerical iterative solutions. Closed-form expression for 3D 3-dB beamwidth in distributed UAV scenarios was derived in~\cite{guo2023impact}, but it is constrained to limited beam pattern and do not support arbitrary beam patterns. In~\cite{miao2021location}, closed-form optimal beamforming weights maximizing the signal to-interference-plus-noise ratio~(SINR) were obtained for 3D UAV scenarios, but the analysis is limited to fixed positioning errors and does not examine performance variations under different error conditions. Overall, there still lacks a systematic analytical framework for characterizing the relationship between beam patterns and communication reliability in an explicit analytical form, which limits the precise optimization of beamforming parameters to meet specified reliability requirements in practical systems.

To provide theoretical guidance for beamforming design, closed-form bounds on outage probability for positioning-assisted beamforming systems were derived for 2D scenarios in~\cite{zhu2022outage}, where the optimal beamwidth is expressed as a function of link distance and transmit power. However, the analysis remains limited to 2D cases and closed-form outage probability expression is not available. Accordingly, the outage probability of 3D positioning-assisted beamforming systems was analyzed in our previous work~\cite{zhang2024positioning}, but the optimal beamwidth analysis remains unaddressed.

In this work, we investigate the positioning-assisted beamforming systems and perform a theoretical analysis of outage probability and optimal beamwidth. The contributions of this work can be summarized as follows:
\begin{itemize}
  \item We derive closed-form expressions of 2D and 3D outage probability, by considering the positioning error distribution, link distance, transmission power and beamwidth. These expressions provide mathematically tractable criterion for beamforming.
  \item The optimal beam patterns are also derived in closed forms for the 2D and 3D cases, which indicates that the beam patterns can be optimized given the transmission power, link distance, and positioning error distribution.
  \item Asymptotic analysis of approximation error is analyzed for the derived outage probability, which quantifies the converging speed of the outage probability expression.
\end{itemize}

The analytical results quantify the impacts of positioning accuracy, link distance, transmission power and beamwidth on the outage probability, and reveal the fact that the beam pattern can be optimized to reduce the outage probability

This paper is organized as follows. In Section~\ref{System Model}, we introduce the system model for both 2D and 3D scenarios. In section~\ref{Closed-Form Outage Probability Analysis}, we derive a closed-form outage probability for the positioning-assisted beamforming system. Section~\ref{Optimization of the Beamwidth} optimizes the beam pattern and introduces the optimal beamwidth expressions. In section~\ref{Asymptotic Error Analysis}, we analyze the approximate error and derive the asymptotic bounds for the outage probability. Section~\ref{Simulation Results} presents numerical results. Finally, Section~\ref{Conclusion} draws some concluding remarks.

\section{System Model}\label{System Model}
\begin{figure}
  \centering
  \includegraphics[trim=2pt 2pt 2pt 2pt,clip,width=0.45\textwidth]{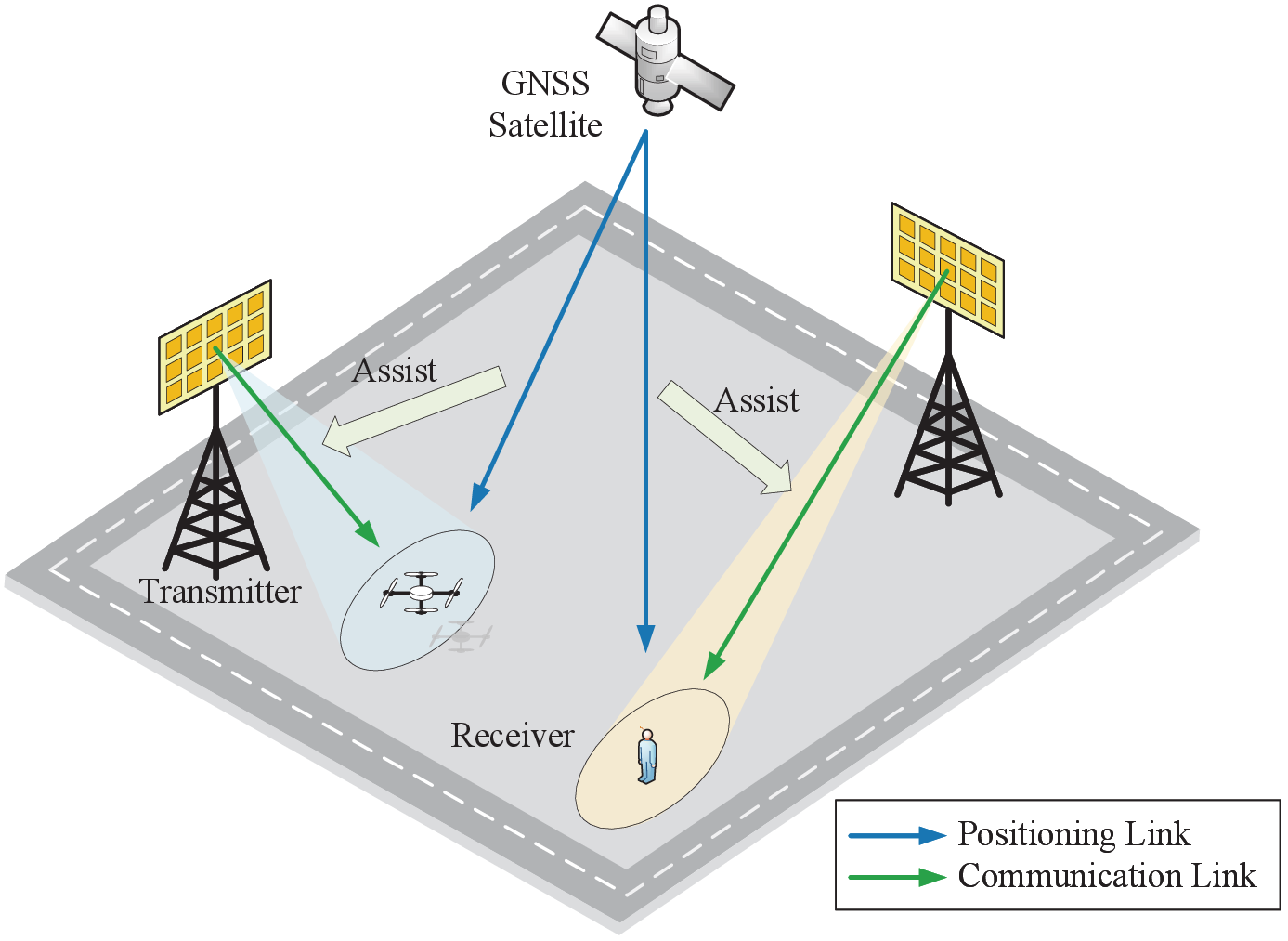}\\
  \caption{The overall scene of positioning-assisted beamforming framework for ISAC systems.}\label{Figure_Overall_Scene}
\end{figure}

The scenario of positioning-assisted beamforming framework is shown in Fig.~\ref{Figure_Overall_Scene}, where the RF transmitter steers the beam towards the receivers, according their positioning results provided by global navigation satellite system~(GNSS), bluetooth low energy~(BLE) or wireless fidelity~(Wi-Fi) systems. The beam pattern can be adjusted by the phase-shifters in the RF chains. In this section, we illustrate the channel model and the beamforming system assisted by the positioning information in both 2D and 3D scenarios, and define the outage probability.

Assuming that the maximum input power of the transmitter antenna is $P_{\max}$, and the received power $P_r$ can be expressed as~\cite{milligan2005modern}
\begin{equation}\label{Eq_Pr}
P_r = P_{\max} A_e G_\theta G_d
\end{equation}
where $A_e$ is the effective area of the receiver antenna and $G_\theta$ is the normalized transmitter antenna gain; $G_d$ represents the free space propagation loss, which can be computed as
\begin{equation}
	G_d = 1 / \left(4 \pi d \right)^2
\end{equation}
where $d$ is the distance between the transmitter antenna and receiver antenna. 

\subsection{2D Beam Pattern and Outage Probability}
In 2D scenarios, $G_\theta$ can be calculated as~\cite{razavizadeh2014three}
\begin{equation}\label{Eq_2D_Gtheta}
	G_\theta = \max \left\{ 10^{-1.2 \frac{\theta^2}{\theta_{3 \text{dB}}^2}}, a_m \right\}
\end{equation}
where $\theta$ is the angle between the beam's boresight direction and receiver direction and $\theta_{3 \text{dB}}$ represents the 3-dB beamwidth of the antenna radiation pattern; $a_m \leq 1$ denotes the slide-lobe level of the antenna pattern, providing a minimum guaranteed propagation gain under low main-lobe gain conditions.

Outage event happens when the received power $P_r$ in~\eqref{Eq_Pr} falls below a threshold $\gamma_{th}$, and thus we can express the 2D outage probability as~\cite{zhu2022outage}
\begin{equation}\label{Eq_2D_Pout_Origin}
\begin{aligned}
P_{\text{out}}^{2\text{D}} \left( \gamma_{th}, \theta_{3 \text{dB}} \right) &\!=\! \Pr  \left( P_r \leq \gamma_{th} \right) \!=\! \Pr \left( P_{\max} A_e G_\theta G_d \leq \gamma_{th} \right) \\
&\!=\! \Pr \left( \! \max \left\{ \! 10^{-1.2 \frac{\theta^2}{\theta_{3 \text{dB}}^2}}, a_m \! \right\} \! \frac{A_e P_{\max}}{\left(4 \pi d \right)^2} \! \leq \! \gamma_{th} \! \right) \!\! .
\end{aligned}
\end{equation}

\begin{figure}
  \centering
  \includegraphics[trim=4pt 1pt 2pt 1pt,clip,width=0.4\textwidth]{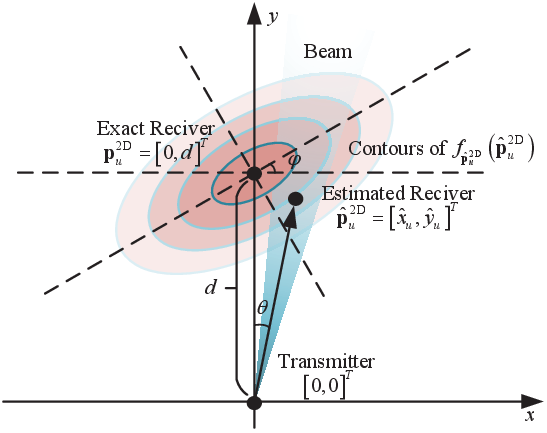}\\
  \caption{Demonstration of a typical positioning-assisted beamforming system. The transmitter is located at the origin, and the exact receiver and estimated received are located at $\mathbf{p}_u^{2\text{D}} = \left[ 0, d \right]^T$ and $\hat{\mathbf{p}}_u^{2\text{D}} = \left[ \hat{x}_u, \hat{y}_u \right]^T$, respectively. The beam points to the estimated receiver position.}\label{Figure_2D_Beam_System_Model}
\end{figure}

Fig.~\ref{Figure_2D_Beam_System_Model} exhibits a typical positioning-assisted beamforming system. Assuming that the exact receiver position is $\mathbf{p}_u^{2\text{D}} = \left[0, d \right]^T$ and the estimated receiver position is $\hat{\mathbf{p}}_u^{2\text{D}} = \left[ \hat{x}_u, \hat{y}_u \right]^T$. $\theta$ represents the error angle and can be computed as
\begin{equation}\label{Eq_2D_theta}
\theta  = \mathrm{atan2} \left( \hat{x}_u, \hat{y}_u \right) \in \left( -\pi, \pi \right]
\end{equation}
where $\mathrm{atan2} \left( \cdot, \cdot \right)$ is the four-quadrant inverse tangent function.

Since $\hat{\mathbf{p}}_u^{2\text{D}}$ follows a bivariate Gaussian distribution, its joint PDF can be expressed as
\begin{equation}
\begin{aligned}
&f_{\hat{\mathbf{p}}_u^{2\text{D}}} \left( \hat{\mathbf{p}}_u^{2\text{D}} \right)  = \frac{1}{2 \pi \left|\mathbf{\Sigma}^{2\text{D}}\right|^{\frac{1}{2}}} \\
& \times  \exp  \left(  -\frac{ \left( \hat{\mathbf{p}}_u^{2\text{D}} - \mathbf{p}_u^{2\text{D}} \right)^{T}  \left( \mathbf{\Sigma}^{2\text{D}} \right)^{-1}  \left( \hat{\mathbf{p}}_u^{2\text{D}} - \mathbf{p}_u^{2\text{D}} \right) }{2} \right)
\end{aligned}
\end{equation}
where $\mathbf{\Sigma}^{2\text{D}} = E \left[ \left( \hat {\mathbf{p}}_u^{2\text{D}} - \mathbf{p}_u^{2\text{D}} \right) \left( \hat {\mathbf{p}}_u^{2\text{D}} - \mathbf{p}_u^{2\text{D}} \right)^T \right] \in \mathbb{R}^{2 \times 2}$ is the 2D covariance matrix of the positioning error vector $\hat {\mathbf{p}}_u^{2\text{D}} - \mathbf{p}_u^{2\text{D}}$ and can be expressed through spectral decomposition as
\begin{equation}
\begin{aligned}
\mathbf{\Sigma}^{2\text{D}} & = \mathbf{R}^{2\text{D}} \mathbf{\Lambda}^{2\text{D}} \left( \mathbf{R}^{2\text{D}}\right)^{T} \\
& =
\begin{bmatrix}
\cos\varphi & -\sin\varphi \\
\sin\varphi & \cos\varphi
\end{bmatrix}
\begin{bmatrix}
\sigma_1^2 & 0 \\
0 & \sigma_2^2
\end{bmatrix}
\begin{bmatrix}
\cos\varphi & \sin\varphi \\
-\sin\varphi & \cos\varphi
\end{bmatrix}
\end{aligned}
\end{equation}
where $\mathbf{R}^{2\text{D}} \in \mathrm{SO}(2)$ is the 2D rotation matrix parameterized by the directional angle $\varphi$; $\sigma_1^2$ and $\sigma_2^2$ denote the variances along the directions of maximum and minimum estimation error, respectively, as shown in Fig.~\ref{Figure_2D_Beam_System_Model}. In this way, we can further rewrite~\eqref{Eq_2D_Pout_Origin} in the following three cases:
\subsubsection{$\gamma_{th} \leq \frac{P_{\max} A_e a_m}{\left( 4 \pi d \right)^2}$}
Since the threshold $\gamma_{th}$ is sufficiently low, the received lobe power $a_m$ always exceeds $\gamma_{th}$, resulting in a zero outage probability, so we have
\begin{equation}
P_{\text{out}}^{2\text{D}} \left( \gamma_{th}, \theta_{3 \text{dB}} \right) = 0.
\end{equation}
\subsubsection{$\gamma_{th} > \frac{P_{\max} A_e}{\left( 4 \pi d \right)^2}$}
The received power falls below the threshold even under perfect pointing, i.e., $\theta = 0$, and thus
\begin{equation}
P_{\text{out}}^{2\text{D}} \left( \gamma_{th}, \theta_{3 \text{dB}} \right) = 1.
\end{equation}
\subsubsection{$\frac{P_{\max} A_e a_m}{\left( 4 \pi d \right)^2} \leq \gamma_{th} < \frac{P_{\max} A_e}{\left( 4 \pi d \right)^2}$}
$\gamma_{th}$ is neither very high or very low, and outage happens when the pointing error is within a specific region. Specifically, $P_{\text{out}}^{2\text{D}} \left( \gamma_{th}, \theta_{3 \text{dB}} \right)$ can be expressed as
\begin{equation}\label{Eq_2D_Pout_Case3_Origin}
	P_{\text{out}}^{2\text{D}} \left( \gamma_{th}, \theta_{3 \text{dB}} \right) = \Pr \left( \theta^2 \geq \frac{\theta_{3 \text{dB}}^2}{1.2}\lg\left( \frac{P_{\max} A_e}{( 4 \pi d )^2\gamma_{th}} \right) \right).
\end{equation}

It is worth noting that~\eqref{Eq_2D_Pout_Case3_Origin} involves integrating a 2D joint Gaussian PDF over a nontrivial region, which cannot yield a closed-form expression. Therefore, an approximation method is to be developed in Section~\ref{Closed-Form Outage Probability Analysis} to enable tractable evaluation of~\eqref{Eq_2D_Pout_Case3_Origin}.

\begin{figure}
	\centering
	\subfloat[3D antenna radiation pattern of $G_{\left( \theta, \phi \right)}$ in~\eqref{Eq_3D_Gtheta} characterized by $\mathbf{A}$.]{
		\includegraphics[width=0.2\textwidth]{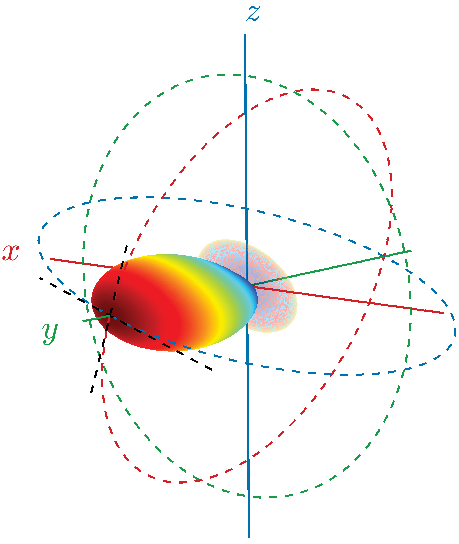}
	}\hfill
	\subfloat[Projection of 3D antenna radiation pattern onto $xoz$ plane.]{
		\includegraphics[width=0.25\textwidth]{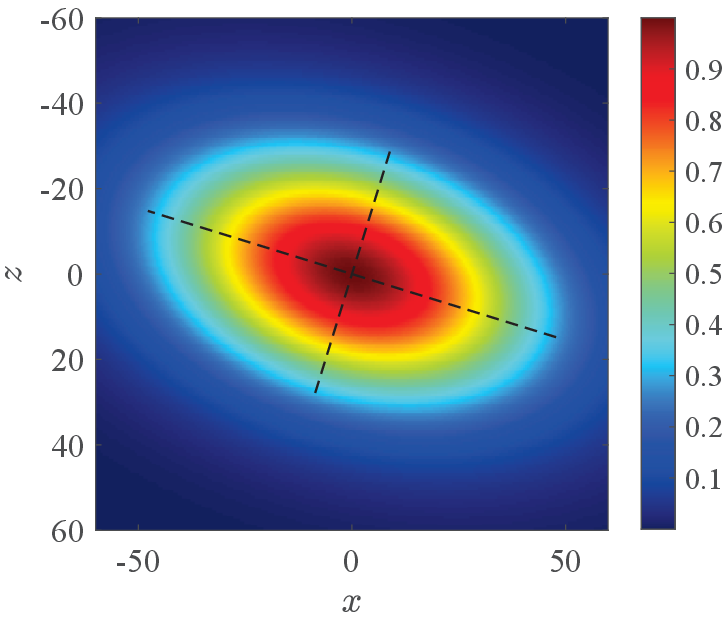}
	}
	\caption{3D antenna radiation pattern of $G_{\left( \theta, \phi \right)}$. The joint Gaussian beam is characterized by $\mathbf{A}$ with $\theta_{3 \text{dB}} = 1.2$~rad, $\phi_{3 \text{dB}} = 0.08$~rad, $\psi = 0.3$~rad.}
	\label{Figure_3D_Antenna_Radiation_Pattern}
\end{figure}

\subsection{3D Beam Pattern and Outage Probability}
In 3D scenarios, we use $\theta$ and $\phi$ to represent the deviations between the boresight and receiving directions in the horizontal and vertical directions, respectively, and the normalized transmitter antenna gain in~\eqref{Eq_2D_Gtheta} should be redefined as~\cite{zhang2024positioning}
\begin{equation}\label{Eq_3D_Gtheta}
G_{\left( \theta, \phi \right)} = \max \left\{ 10^{-1.2 \left[ \theta, \phi \right] \mathbf{A} \left[ \theta, \phi \right]^T}, a_m \right\}
\end{equation}
where
\begin{equation}\label{Eq_3D_A}
\mathbf{A} = 
\begin{bmatrix}
\frac{1}{\theta_{3 \text{dB}}^2} & m \\
m & \frac{1}{\phi_{3 \text{dB}}^2}
\end{bmatrix}
\end{equation}
represents the 3D beam pattern in terms of its principal axes and orientation, where $\theta_{3 \text{dB}}$ and $\phi_{3 \text{dB}}$ represent the horizontal and vertical 3-dB beamwidth of the antenna radiation pattern, and $m$ represents the beam rotation parameterized by beam rotation angle $\psi$. The antenna radiation pattern of $G_{\left( \theta, \phi \right)}$ in~\eqref{Eq_3D_Gtheta} characterized by $\mathbf{A}$ and its projection onto $xoz$ plane are shown in Fig.~\ref{Figure_3D_Antenna_Radiation_Pattern}. Hence, the outage probability of 3D beamforming assisted by positioning information is rewritten as
\begin{equation}\label{Eq_3D_Pout_Origin}
\begin{aligned}
& P_{\text{out}}^{3\text{D}} \left( \gamma_{th}, \mathbf{A} \right) = \Pr  \left( P_r \leq \gamma_{th} \right) \\
& = \Pr  \left( P_{\max} A_e G_{\left( \theta, \phi \right)} G_d \leq \gamma_{th} \right) \\
& = \Pr \left( \max  \left\{ 10^{-1.2 \left[ \theta, \phi \right] \mathbf{A} \left[ \theta, \phi \right]^T} , a_m \right\}  \frac{P_{\max} A_e}{\left(4 \pi d \right)^2}  \leq  \gamma_{th}  \right) .
\end{aligned}
\end{equation}

\begin{figure}
	\centering
	\includegraphics[trim=4pt 1pt 2pt 1pt,clip,width=0.4\textwidth]{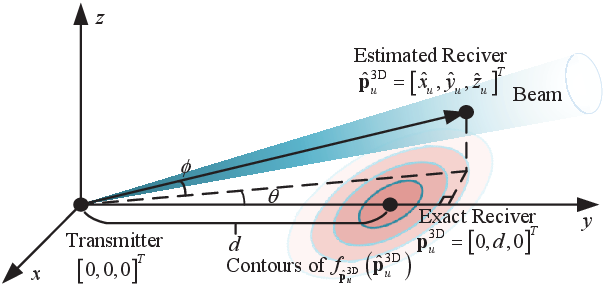}\\
	\caption{Demonstration of a general 3D positioning-assisted beamforming system. The transmitter is located at the origin, and the exact receiver and estimated received are located at $\mathbf{p}_u^{3\text{D}} = \left[ 0, d, 0 \right]^T$ and $\hat{\mathbf{p}}_u^{3\text{D}} = \left[ \hat{x}_u, \hat{y}_u, \hat{z}_u \right]^T$, respectively. The beam points to the estimated receiver position.}\label{Figure_3D_Beam_System_Model}
\end{figure}

A general 3D positioning-assisted beamforming system is shown in Fig.~\ref{Figure_3D_Beam_System_Model}. The exact receiver is located at $\mathbf{p}_u^{3\text{D}} = \left[ 0, d, 0 \right]^T$ and the estimated receiver position is $\hat{\mathbf{p}}_u^{3\text{D}} = \left[ \hat{x}_u, \hat{y}_u, \hat{z}_u \right]^T$. The transmitter is located at the origin and we can compute $\theta$ and $\phi$ as
\begin{equation}\label{Eq_3D_theta}
\theta  = \mathrm{atan2} \left( \hat{x}_u, \hat{y}_u \right) \in \left( -\pi, \pi \right]
\end{equation}
and
\begin{equation}\label{Eq_3D_phi}
\phi = \arctan\frac{\hat{z}_u}{\sqrt{\hat{x}_u^2 + \hat{y}_u^2}} \in \left( -\pi / 2, \pi / 2 \right].
\end{equation}

Similar to the 2D scenarios, $\hat{\mathbf{p}}_u^{3\text{D}}$ follows a trivariate Gaussian distribution with its PDF expressed as
\begin{equation}
\begin{aligned}
&f_{\hat{\mathbf{p}}_u^{3\text{D}}} \left( \hat{\mathbf{p}}_u^{3\text{D}} \right) = \frac{1}{\left( 2 \pi \right)^{\frac{3}{2}} \left|\mathbf{\Sigma}^{3\text{D}}\right|^{\frac{1}{2}}} \\
& \times \exp \left( -\frac{ \left( \hat{\mathbf{p}}_u^{3\text{D}} - \mathbf{p}_u^{3\text{D}} \right)^T \left( \mathbf{\Sigma}^{3\text{D}} \right)^{-1} \left( \hat{\mathbf{p}}_u^{3\text{D}} - \mathbf{p}_u^{3\text{D}} \right) }{2} \right)
\end{aligned}
\end{equation}
where $\mathbf{\Sigma}^{3\text{D}} = E \left[ \left( \hat {\mathbf{p}}_u^{3\text{D}} - \mathbf{p}_u^{3\text{D}} \right) \left( \hat {\mathbf{p}}_u^{3\text{D}} - \mathbf{p}_u^{3\text{D}} \right)^T \right] \in \mathbb{R}^{3 \times 3}$ represents the 3D covariance matrix of the positioning error vector $\hat{\mathbf{p}}_u^{3\text{D}} - \mathbf{p}_u^{3\text{D}}$ and can be eigendecomposed  as
\begin{equation}
\mathbf{\Sigma}^{3\text{D}} = \mathbf{R}^{3\text{D}} \mathbf{\Lambda}^{3\text{D}} \left( \mathbf{R}^{3\text{D}} \right)^T
\end{equation}
where $\mathbf{R}^{3\text{D}} \in \mathrm{SO}(3)$ represents the 3D rotation matrix parameterized by directional angle $\varphi_x, \varphi_y, \varphi_z$ and $\mathbf{\Lambda}^{3\text{D}} = \mathrm{diag}(\sigma_1^2, \sigma_2^2, \sigma_3^2)$ is a diagonal matrix of eigenvalues satisfying $\sigma_1^2 > \sigma_2^2 > \sigma_3^2$. Hence, eq.~\eqref{Eq_3D_Pout_Origin} can be rewritten as follows:
\subsubsection{$\gamma_{th} \leq \frac{P_{\max} A_e a_m}{\left( 4 \pi d \right)^2}$}
Since the threshold $\gamma_{th}$ is sufficiently low, the received lobe power $a_m$ always exceeds $\gamma_{th}$, resulting in a zero outage probability, so we have
\begin{equation}
P_{\text{out}}^{3\text{D}} \left( \gamma_{th}, \mathbf{A} \right) = 0.
\end{equation}
\subsubsection{$\gamma_{th} > \frac{P_{\max} A_e}{\left( 4 \pi d \right)^2}$}
The received power falls below the threshold even under perfect pointing, i.e., $\theta = 0, \phi = 0$, and thus
\begin{equation}
P_{\text{out}}^{3\text{D}} \left( \gamma_{th}, \mathbf{A} \right) = 1.
\end{equation}
\subsubsection{$\frac{P_{\max} A_e a_m}{\left( 4 \pi d \right)^2} \leq \gamma_{th} < \frac{P_{\max} A_e}{\left( 4 \pi d \right)^2}$}
$\gamma_{th}$ is neither too high or too low, and outage happens when the pointing error is within a specific region. Specifically, $P_{\text{out}}^{3\text{D}} \left( \gamma_{th}, \mathbf{A} \right)$ can be expressed as
\begin{equation}\label{Eq_3D_Pout_Case3_Origin}
P_{\text{out}}^{3\text{D}} \left( \gamma_{th}, \mathbf{A} \right) = \Pr \left( \left[ \theta, \phi \right] \mathbf{A} \left[ \theta, \phi \right]^T \geq  \frac{1}{1.2} \lg \frac{P_{\max} A_e}{\left( 4 \pi d \right)^2 \gamma_{th}} \right)  .
\end{equation}

\begin{figure}
	\centering
	\includegraphics[trim=4pt 2pt 2pt 2pt,clip,width=0.4\textwidth]{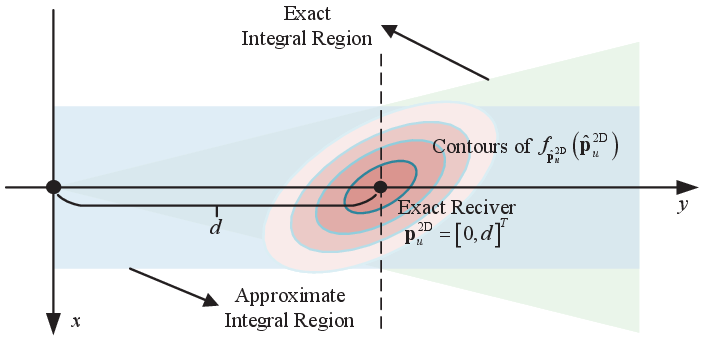}\\
	\caption{Equivalent model of positioning-assisted beamforming system. Triangular and rectangular region represent exact and approximate integral region, respectively.}
	\label{Figure_2D_Integral_Region}
\end{figure}

Similar to~\eqref{Eq_2D_Pout_Case3_Origin}, eq.~\eqref{Eq_3D_Pout_Case3_Origin}  also characterizes the integral of a 2D normal distribution over a nontrivial region, where deriving a closed-form analytical expression is generally difficult. Consequently, Section~\ref{Closed-Form Outage Probability Analysis} introduces an asymptotically tight approximation for~\eqref{Eq_3D_Pout_Case3_Origin}.

\section{Asymptotic Outage Probability Analysis}\label{Closed-Form Outage Probability Analysis}
In this section, we simplify the outage probability expressions in~\eqref{Eq_2D_Pout_Case3_Origin} for the 2D scenario and~\eqref{Eq_3D_Pout_Case3_Origin} for the 3D scenario.

\subsection{2D Asymptotic Outage Probability}
Substituting~\eqref{Eq_2D_theta} into~\eqref{Eq_2D_Pout_Case3_Origin}, we can rewrite~\eqref{Eq_2D_Pout_Case3_Origin} as
\begin{equation}\label{Eq_2D_Pout_Case3_New}
\begin{aligned}
P_{\text{out}}^{2\text{D}} \left( \gamma_{th}, \theta_{3 \text{dB}} \right) &= \! \Pr \left( \theta^2 \geq \frac{\theta_{3 \text{dB}}^2}{1.2} \lg \left( \frac{P_{\max} A_e}{( 4 \pi d )^2\gamma_{th}} \right) \right) \\
&\approx \! \Pr \left( \frac{ \left|\hat{x}_u\right| }{\hat{y}_u} \geq \tan\sqrt{\frac{\theta_{3 \text{dB}}^2}{1.2}\lg\left( \frac{P_{\max} A_e}{( 4 \pi d )^2\gamma_{th}} \right)}  \right) \\
&= \Pr \left( \left|\hat{x}_u\right| \geq k \hat{y}_u  \right) \\
&=\! \iint_{\left| x \right| \geq k y} f_{\hat{\mathbf{p}}_u^{2\text{D}}} \left(\hat{\mathbf{p}}_u^{2\text{D}} \right) dx dy
\end{aligned}
\end{equation}
where
\begin{equation}\label{Eq_2D_k}
k = \tan\sqrt{\frac{\theta_{3 \text{dB}}^2}{1.2}\lg\left( \frac{P_{\max} A_e}{(4\pi d )^2\gamma_{th}} \right) }.
\end{equation}
Given that the positioning error is sufficiently smaller than the link distance, we have $\hat{y}_u / d \rightarrow 1$ and \eqref{Eq_2D_Pout_Case3_New} can be approximated by
\begin{equation}\label{Eq_2D_Pout_Approximate}
\begin{aligned}
&P_{\text{out}}^{2\text{D}, \infty} \left( \gamma_{th}, \theta_{3 \text{dB}} \right) \\
&\approx  \Pr \left(\left|\hat{x}_u\right| \geq k \hat{y}_u \right) \approx \Pr \left(\left|\hat{x}_u \right| \geq k d \right)\\
&= 1 - \int_{-k d}^{k d} \left( \int_{-\infty}^{+\infty} f_{\hat{y}_u} \left( y \right) \, dy \right) \cdot f_{\hat{x}_u} \left( x \right) \, dx \\
&= 1 - \int_{-k d}^{k d} f_{\hat{x}_u} \left( x \right)\, d x \\
& = 2Q \left( \frac{d}{\sigma_x} \tan \sqrt{\frac{\theta_{3 \text{dB}}^2}{1.2}\lg\left( \frac{P_{\max} A_e}{\left( 4 \pi d \right)^2\gamma_{th}} \right)} \right)
\end{aligned}
\end{equation}
where $Q(x) = \int_x^{+\infty} \frac{1}{\sqrt {2 \pi}} \exp \left( -t^2 / 2 \right) \, dt $ denotes the Gaussian $Q$-function and $\sigma_x = \sqrt{\sigma_1^2 \cos^2\varphi + \sigma_2^2 \sin^2\varphi}$ is the standard deviation of $\hat{x}_u$. The exact integral region in~\eqref{Eq_2D_Pout_Case3_Origin} and the approximate integral region in~\eqref{Eq_2D_Pout_Approximate} is shown in Fig.~\ref{Figure_2D_Integral_Region}.

\subsection{3D Asymptotic Outage Probability}

\begin{figure}
  \centering
  \includegraphics[trim=4pt 1pt 2pt 1pt,clip,width=0.4\textwidth]{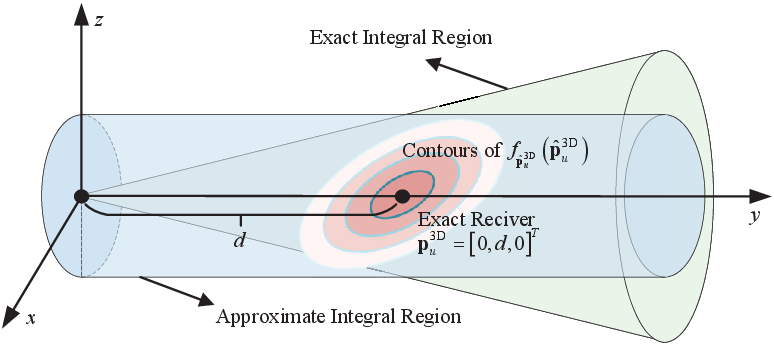}\\
  \caption{Equivalent model of 3D positioning-assisted beamforming system. Conical and cylindrical region represent exact and approximate integral region, respectively.}\label{Figure_3D_Integral_Region}
\end{figure}

As the components $\hat{x}_u$ and $\hat{z}_u$ are contributed by the positioning errors and they are incomparable to the link distance, we have $\hat{y}_u / d \rightarrow 1$. To facilitate further analysis, we employ Taylor series expansions of~\eqref{Eq_3D_theta} and~\eqref{Eq_3D_phi} to obtain analytical approximations as
\begin{equation}\label{Eq_3D_theta_Taylor}
\theta =  \mathrm{atan2} \left( \hat{x}_u, \hat{y}_u \right) = \frac{\hat{x}_u}{d} + o \left( \frac{\hat{x}_u}{d} \right)
\end{equation}
and
\begin{equation}\label{Eq_3D_phi_Taylor}
\phi = \arctan\frac{\hat{z}_u}{\sqrt{\hat{x}_u^2 + \hat{y}_u^2}} = \frac{\hat{z}_u}{d} + o \left( \frac{\hat{z}_u}{d} \right).
\end{equation}
Substituting~\eqref{Eq_3D_theta_Taylor} and~\eqref{Eq_3D_phi_Taylor} into~\eqref{Eq_3D_Pout_Case3_Origin}, eq.~\eqref{Eq_3D_Pout_Case3_Origin} can be expressed approximately as
\begin{equation}\label{Eq_3D_Pout_Case3_New}
\begin{aligned}
& P_{\text{out}}^{3\text{D}, \infty} \left( \gamma_{th}, \mathbf{A} \right) \\
&= \Pr \left( \left[ \hat{x}_u, \hat{z}_u \right] \mathbf{A} \left[ \hat{x}_u, \hat{z}_u \right]^T \geq \frac{d^2}{1.2} \lg \frac{P_{\max} A_e}{\left( 4 \pi d \right)^2 \gamma_{th}} \right)
\end{aligned}
\end{equation}
where the approximation procedure is illustrated in Fig.~\ref{Figure_3D_Integral_Region}
The joint PDF of $\left[ \hat{x}_u, \hat{z}_u \right]$ in~\eqref{Eq_3D_Pout_Case3_New} can be calculated by marginalizing $\hat y_u$ as
\begin{equation}\label{Eq_3D_PDF_xz}
\begin{aligned}
f_{\hat{x}_u,\hat{z}_u} \left( x,z \right) &= \int_{-\infty}^{+\infty} f_{\hat{\mathbf{p}}_u^{3\text{D}}} \left( \hat{\mathbf{p}}_u^{3\text{D}} \right)  dy \\
&= \frac{1}{2 \pi \!\left| \mathbf{\Sigma}^{'2\text{D}} \right|^{\!\frac{1}{2}}} 
 \!\exp \!\left( \!-\frac{ \left[ x, z \right]\!
	\left( \mathbf{\Sigma}^{'2\text{D}} \right)^{-1}
	\left[ x, z \right]^T }{2} \right) \! .
\end{aligned}
\end{equation}
where
\begin{equation}\label{Eq_3D_Cov_Matrix_xz}
\mathbf{\Sigma}^{'2\text{D}} = 
 \begin{bmatrix}
\mathbf{\Sigma}^{3\text{D}} \left( 1,1 \right) &\mathbf{\Sigma}^{3\text{D}} \left( 1,3 \right) \\ \mathbf{\Sigma}^{3\text{D}} \left( 3,1 \right)  & \mathbf{\Sigma}^{3\text{D}} \left( 3,3 \right)
\end{bmatrix}
\end{equation}
is the covariance matrix of $\left[ \hat{x}_u, \hat{z}_u \right]$ and $\mathbf{\Sigma}^{3\text{D}} \left( m,n \right)$ represents the $\left( m,n \right)$th entry of $\mathbf{\Sigma}^{3\text{D}}$. Combining~\eqref{Eq_3D_Pout_Case3_New},~\eqref{Eq_3D_PDF_xz} and~\eqref{Eq_3D_Cov_Matrix_xz}, we can calculate~\eqref{Eq_3D_Pout_Case3_New} as 
\begin{equation}\label{Eq_3D_Pout_To_Be_Opt}
\begin{aligned}
P_{\text{out}}^{3\text{D}, \infty} \left( \gamma_{th}, \mathbf{A} \right) &= \iint\limits_{\substack{\left[ x, z \right] \mathbf{A} \left[ x, z \right]^T \\ \geq \frac{d^2}{1.2} \lg \frac{P_{\max} A_e}{\left( 4 \pi d \right)^2 \gamma_{th}}}}f_{\hat{x}_u,\hat{z}_u} \left( x,z \right) dx dz \\
&= \iint\limits_{\substack{\left[ x, z \right] \mathbf{A} \left[ x, z \right]^T \\ \geq \frac{d^2}{1.2} \lg \frac{P_{\max} A_e}{\left( 4 \pi d \right)^2 \gamma_{th}}}}
\frac{1}{2 \pi \left| \mathbf{\Sigma}^{'2\text{D}} \right|^{\frac{1}{2}}} \\
&\quad \times \exp \! \left( \! -\frac{ \left[ x, z \right]
\left( \mathbf{\Sigma}^{'2\text{D}} \! \right)^{-1}
\left[ x, z \right]^T }{2} \right) \! dx  dz.
\end{aligned}
\end{equation}
Applying the eigenvalue decomposition of $\mathbf{A}$ as $\mathbf{A} = \mathbf{U} \mathbf{Y} \mathbf{U}^T$ and defining $\mathbf{t} = \sqrt{\mathbf{Y}} \mathbf{U}^T \left[ x, z \right]^T$, we can transform~\eqref{Eq_3D_Pout_To_Be_Opt} to
\begin{equation}\label{Eq_3D_Pout_t}
\begin{aligned}
&P_{\text{out}}^{3\text{D}, \infty} \left( \gamma_{th}, \mathbf{A} \right) \\
&= \iint\limits_{\mathbf{t}^T \mathbf{t} \geq \frac{d^2}{1.2} \lg \frac{P_{\max} A_e}{\left( 4 \pi d \right)^2 \gamma_{th}}}
\frac{1}{2 \pi \left| \mathbf{Y} \mathbf{\Sigma}^{'2\text{D}} \right|^{\frac{1}{2}}} \\
&\quad \times \exp \left( -\frac{\mathbf{t}^T \sqrt{\mathbf{Y}^{-1}} \mathbf{U}^T \left( \mathbf{\Sigma}^{'2\text{D}} \right)^{-1} \mathbf{U} \sqrt{\mathbf{Y}^{-1}} \mathbf{t}}{2} \right)
\, d\mathbf{t}.
\end{aligned}
\end{equation}
Let $\mathbf{H}^{-1} = \sqrt{\mathbf{Y}^{-1}} \mathbf{U}^T \left( \mathbf{\Sigma}^{'2\text{D}} \right)^{-1} \mathbf{U} \sqrt{\mathbf{Y}^{-1}}$  and the eigenvalue decomposition of $\mathbf{H}$ is given by
$\mathbf{H} = \mathbf{V} \mathbf{D} \mathbf{V}^T$, where $\mathbf{D} = \mathrm{diag} \left( \lambda_1^2, \lambda_2^2 \right)$ and $\lambda_1 \leq \lambda_2$. Define $\mathbf{q} = \mathbf{V}^T \mathbf{t}$ and~\eqref{Eq_3D_Pout_t} can be further calculated as
\begin{equation}\label{Eq_3D_Pout_q}
\begin{aligned}
&P_{\text{out}}^{3\text{D}, \infty} \left( \gamma_{th}, \mathbf{A} \right) \\
& = \iint\limits_{\mathbf{q}^T \mathbf{q} \geq \frac{d^2}{1.2} \lg \frac{P_{\max} A_e}{\left( 4 \pi d \right)^2 \gamma_{th}}}
\frac{1}{2 \pi \left| \mathbf{D} \right|^{\frac{1}{2}}}\exp \left( -\frac{\mathbf{q}^T \mathbf{D}^{-1} \mathbf{q}}{2} \right)
\, d\mathbf{q}
\end{aligned}
\end{equation}
where $\mathbf{q} = \left( q_1, q_2 \right)$. Furthermore, eq.~\eqref{Eq_3D_Pout_q} can be categorized into the following two cases, depending on whether $\lambda_1$ and $\lambda_2$ are equal.
\subsubsection{$\lambda_1 < \lambda_2$}
Define $q= \lambda_1 / \lambda_2 \in \left( 0,1 \right)$ and $\omega = \lambda_1^2 + \lambda_2^2$~\cite{jajszczyk2001digital}, and \eqref{Eq_3D_Pout_q} can be expressed as~\cite{zhu2018new}
\begin{equation}\label{Eq_3D_Pout_hoyt}
P_{\text{out}}^{3\text{D}, \infty} \left( \gamma_{th}, \mathbf{A} \right) = 1 - \int_0^{\sqrt{\frac{d^2}{1.2} \lg \frac{P_{\max} A_e}{\left( 4 \pi d \right)^2 \gamma_{th}}}} h\left( x; q, \omega \right) \, dx
\end{equation}
where $h\left( x; q, \omega \right)$ is the PDF of Hoyt distribution and its cumulative distribution function~(CDF) can be calculated as~\cite{lopez2015asymptotically}
\begin{equation}\label{Eq_CDF_Hoyt}
\begin{aligned}
H \left( x; q, \omega \right) &= \int_0^x h\left( x; q, \omega \right) \, dx \\
&= Q_1 \left( \frac{x \sqrt{1 - q^4}}{2 q \sqrt{\omega}} \sqrt{\frac{1 + q}{1 - q}}, \frac{x \sqrt{1 - q^4}}{2 q \sqrt{\omega}} \sqrt{\frac{1 - q}{1 + q}} \right) \\
&- Q_1 \left( \frac{x \sqrt{1 - q^4}}{2 q \sqrt{\omega}} \sqrt{\frac{1 - q}{1 + q}}, \frac{x \sqrt{1 - q^4}}{2 q \sqrt{\omega}} \sqrt{\frac{1 + q}{1 - q}} \right)
\end{aligned}
\end{equation}
where $Q_1 \left( a, b \right) = \int_b^{+\infty} x \exp \left( - \frac{x^2 + a^2}{2} \right) I_0 \left( a x \right) \, dx$ is the first order Marcum $Q$-function; $I_0 \left( x \right)$ is the first order modified Bessel function of first kind. Substituting~\eqref{Eq_CDF_Hoyt} into~\eqref{Eq_3D_Pout_hoyt}, we have
\begin{equation}\label{Eq_3D_Pout_lambda _unequal}
\begin{aligned}
&P_{\text{out}}^{3\text{D}, \infty} \left( \gamma_{th}, \mathbf{A} \right) \\
&= 1 -  Q_1 \left( \frac{\sqrt{\left( 1 - q^4 \right) \frac{d^2}{1.2} \lg \frac{P_{\max} A_e}{\left( 4 \pi d \right)^2 \gamma_{th}}}}{2 q \sqrt{\omega}} \sqrt{\frac{1 + q}{1 - q}}, \right. \\
& \qquad \qquad \quad \left. \frac{\sqrt{\left( 1 - q^4 \right) \frac{d^2}{1.2} \lg \frac{P_{\max} A_e}{\left( 4 \pi d \right)^2 \gamma_{th}}}}{2 q \sqrt{\omega}} \sqrt{\frac{1 - q}{1 + q}} \right) \\
& \quad+  Q_1 \left( \frac{\sqrt{\left( 1 - q^4 \right) \frac{d^2}{1.2} \lg \frac{P_{\max} A_e}{\left( 4 \pi d \right)^2 \gamma_{th}}}}{2 q \sqrt{\omega}} \sqrt{\frac{1 - q}{1 + q}}, \right. \\
& \qquad \qquad \quad \left. \frac{\sqrt{\left( 1 - q^4 \right) \frac{d^2}{1.2} \lg \frac{P_{\max} A_e}{\left( 4 \pi d \right)^2 \gamma_{th}}}}{2 q \sqrt{\omega}} \sqrt{\frac{1 + q}{1 - q}} \right).
\end{aligned}
\end{equation}
\subsubsection{$\lambda_1 = \lambda_2$}
Define $\lambda = \lambda_1 = \lambda_2$, we can calculate~\eqref{Eq_3D_Pout_q} as
\begin{equation}\label{Eq_3D_Pout_lambda _equal}
\begin{aligned}
&P_{\text{out}}^{3\text{D}, \infty} \left( \gamma_{th}, \mathbf{A} \right) \\
&= \iint\limits_{q_1^2 + q_2^2 \geq \frac{d^2}{1.2} \lg \frac{P_{\max} A_e}{\left( 4 \pi d \right)^2 \gamma_{th}}} \frac{1}{2 \pi \lambda^2}
\exp \left( -\frac{q_1^2 + q_2^2}{2 \lambda^2} \right) \, dq_1 \, dq_2 \\
& = \exp \left( -\frac{d^2}{2.4 \lambda^2} \lg \frac{P_{\max} A_e}{\left( 4 \pi d \right)^2 \gamma_{th}} \right).
\end{aligned}
\end{equation}

It can be observed from~\eqref{Eq_3D_Pout_lambda _equal} that the outage probability admits a compact closed-form expression in terms of exponential functions when $\lambda_1 = \lambda_2$, eliminating the need for Marcum $Q$-function. This is because the Hoyt distribution PDF in~\eqref{Eq_3D_Pout_hoyt} specialized to the Rayleigh distribution PDF, which is consistent with the expression reported in~\cite{zhang2024positioning}.

\section{Optimization of the Beam Pattern}\label{Optimization of the Beamwidth}
Since we have derived the closed-form expression of the outage probability, it is ready to further optimize the beamwidth for a lower outage probability. Specifically, we optimize the beam pattern in order to minimize the expressions $P_{\text{out}}^{2\text{D}} \left( \gamma_{th}, \theta_{3 \text{dB}} \right)$.
\subsection{2D Optimal Beam Pattern}
Assuming that the transmit power $P_t$ is fixed and the side-lobe level $a_m$ is negligible, $P_t$ can be computed as
\begin{equation}\label{Eq_2D_Pt}
\begin{aligned}
P_t &= \int_{-\pi}^{\pi} P_{\max} G_\theta \, d\theta \\
&= \frac{P_{\max} \theta_{3\text{dB}} \sqrt\pi}{\sqrt{1.2 \ln10}} \left[ 1 - 2Q\left( \frac{\pi \sqrt{2.4 \ln10}}{\theta_{3\text{dB}}} \right) \right].
\end{aligned}
\end{equation}
Noting that $Q\left( \frac{\pi \sqrt{2.4 \ln 10}}{\theta_{3\text{dB}}} \right) \approx 0$ when $\theta_{3\text{dB}} < \pi$, we can approximate~\cite{zhu2022outage}
\begin{equation}
P_t   \approx P_{\max} \frac{\theta_{3\text{dB}} \sqrt\pi}{\sqrt{1.2 \ln10}}
\end{equation}
and
\begin{equation}\label{Eq_2D_Pmax}
P_{\max} \approx P_t \frac{\sqrt{1.2 \ln10}}{\theta_{3\text{dB}} \sqrt\pi}.
\end{equation}

Substituting~\eqref{Eq_2D_Pmax} into~\eqref{Eq_2D_k}, we minimize the outage probability by maximizing $k$ in~\eqref{Eq_2D_k} as
\begin{equation}\label{Eq_2D_Optimizer}
\max_{\theta_{3 \text{dB}}} \theta_{3 \text{dB}}^2 \lg \left( \frac{A_e P_t \sqrt{1.2 \ln10}}{\theta_{3\text{dB}} \left( 4 \pi d \right)^2\gamma_{th} \sqrt\pi} \right).
\end{equation}
Taking the derivative of~\eqref{Eq_2D_Optimizer} and setting it to zero, the maximizer of~\eqref{Eq_2D_Optimizer} is obtained as
\begin{equation}\label{Eq_2D_Optimal theta}
\theta_{3 \text{dB}}^\ast = \frac{A_e P_t}{\left( 4 \pi d \right)^2\gamma_{th}} \sqrt{\frac{1.2 \ln10}{e \pi}}.
\end{equation}
Noting that the second-order derivative of the objective function in~\eqref{Eq_2D_Optimizer} is $2 \lg \left( \frac{A_e P_t \sqrt{1.2 \ln10}}{\theta_{3\text{dB}} \left( 4 \pi d \right)^2\gamma_{th} \sqrt\pi} \right) - \frac{3}{\ln10} > 0$ and applying~\eqref{Eq_2D_Optimal theta} to~\eqref{Eq_2D_k}, we have the global outage probability minimizer
\begin{equation}\label{Eq_2D_Optimal_k}
k^\ast = \tan \left( {\frac{A_e P_t}{\left( 4 \pi d \right)^2\gamma_{th}} \sqrt{\frac{1}{2 e \pi}}} \right)
\end{equation}
and the lowest outage probability in~\eqref{Eq_2D_Pout_Approximate} is
\begin{equation}\label{Eq_2D_Pout_Optimal}
P_{\text{out}}^{2\text{D}, \ast} = 2Q \left( \frac{d}{\sigma_x} \tan \left( {\frac{A_e P_t}{\left( 4 \pi d \right)^2\gamma_{th}} \sqrt{\frac{1}{2 e \pi}}} \right) \right).
\end{equation}

Equation~\eqref{Eq_2D_Optimal theta} provides a closed-form expression for the optimal beamwidth as a function of the transmit power $P_t$, the link distance $d$, and the outage threshold $\gamma_{th}$. This expression enables direct evaluation of the optimal beam configuration without relying on numerical optimization or exhaustive search. As $d$ increases, the optimal beamwidth decreases accordingly, producing a narrower beam that helps counteract the increased pathloss and concentrate energy more effectively along the intended direction. Conversely, when $P_t$ increases, a wider beamwidth is preferred to encompass a broader angular region, covering a larger region of potential user distribution and enhancing the system's robustness against beam misalignment caused by positioning errors or angular uncertainties. This trade-off between directional focus and angular coverage reflects the fundamental design considerations in directional communication systems and highlights the practical value of the derived expression in~\eqref{Eq_2D_Optimal_k}.

It is worth noting that the optimization carried out in this work is performed on the approximate outage probability in~\eqref{Eq_2D_Pout_Approximate}, while the study in~\cite{zhu2022outage} optimizes the exact  outage probability in~\eqref{Eq_2D_Pout_Case3_New}. Despite the difference in the objective functions, the conclusion in~\eqref{Eq_2D_Optimal_k} is consistent with that reported in~\cite{zhu2022outage}.

\subsection{3D Optimal Beam Pattern}
We start from the outage probability expression in~\eqref{Eq_3D_Pout_Case3_New} and aim to minimize it via beam pattern optimization. Since the random variables $\hat{x}_u,\hat{z}_u$ in~\eqref{Eq_3D_Pout_Case3_New} do not follow a standard Gaussian distribution, the beam pattern optimization problem becomes analytically intractable. Therefore, we first transform them into an equivalent standard Gaussian form.

First, we apply the eigenvalue decomposition of the covariance matrix of $\left[ \hat{x}_u, \hat{z}_u \right]$ in~\eqref{Eq_3D_Cov_Matrix_xz} as $\mathbf{\Sigma}^{'2\text{D}} = \mathbf{G} \mathbf{\Omega} \mathbf{G}^T$. Then, we define $\mathbf{w} = \sqrt{\mathbf{\Omega}^{-1}} \mathbf{G}^T \left[ x, z \right]^T$ and~\eqref{Eq_3D_Pout_Case3_New} can be reformulated as
\begin{equation}\label{Eq_3D_Pout_Case3_w}
\begin{aligned}
& P_{\text{out}}^{3\text{D}, \infty} \left( \gamma_{th}, \mathbf{A} \right) \\
&= \Pr \left( \mathbf{w}^T \sqrt{\mathbf{\Omega}} \mathbf{G}^T \mathbf{A} \mathbf{G} \sqrt{\mathbf{\Omega}} \mathbf{w} \geq \frac{d^2}{1.2} \lg \frac{P_{\max} A_e}{\left( 4 \pi d \right)^2 \gamma_{th}} \right).
\end{aligned}
\end{equation}
Define 
\begin{equation}\label{Eq_3D_B}
\mathbf{B} = \sqrt{\mathbf{\Omega}} \mathbf{G}^T \mathbf{A} \mathbf{G} \sqrt{\mathbf{\Omega}},
\end{equation}
and perform its eigenvalue decomposition as
\begin{equation}\label{Eq_3D_B_Eigen}
	\mathbf{B} = \mathbf{Q} \mathbf{M} \mathbf{Q}^T
\end{equation}
where $\mathbf{M} = \mathrm{diag} \left( \xi_1, \xi_2 \right)$ is a diagonal matrix, and $\xi_1, \xi_2$ are eigenvalues of $\mathbf{B}$. Let $\mathbf{v} = \mathbf{Q}^T \mathbf{w}$ and~\eqref{Eq_3D_Pout_Case3_w} can be reformulated as
\begin{equation}\label{Eq_3D_Pout_Case3_v}
\begin{aligned}
P_{\text{out}}^{3\text{D}, \infty} \left( \gamma_{th}, \mathbf{A} \right) &= \Pr \left( \mathbf{v}^T  \mathbf{M}  \mathbf{v} \geq \frac{d^2}{1.2} \lg \frac{P_{\max} A_e}{\left( 4 \pi d \right)^2 \gamma_{th}} \right) \\
& =\Pr \left( \xi_1 v_1^2 + \xi_2 v_2^2  \geq \frac{d^2}{1.2} \lg \frac{P_{\max} A_e}{\left( 4 \pi d \right)^2 \gamma_{th}} \right)
\end{aligned}
\end{equation}
where $\mathbf{v} = \left( v_1, v_2 \right)$ and $v_1, v_2 \sim \mathcal{N} \left( 0, 1 \right)$ are independent. As shown in~\eqref{Eq_3D_Pout_Case3_v}, the outage probability expression in~\eqref{Eq_3D_Pout_Case3_New} is reformulated in terms of standard Gaussian random variables $v_1, v_2$, which significantly facilitates the subsequent analytical derivations.

Since $P_{\max}$ in~\eqref{Eq_3D_Pout_Case3_v} is a function of  $\mathbf{A}$, which in turn depends on the beam pattern parameters $\theta_{3\text{dB}}$ and $\phi_{3\text{dB}}$ to be optimized, $P_{\max}$ must be explicitly evaluated first. Similar to~\eqref{Eq_2D_Pt}, the transmit power in 3D scenarios can be computed as
\begin{equation}\label{Eq_3D_Pt}
\begin{aligned}
P_t &= \int_{-\pi / 2}^{\pi / 2} \int_{-\pi}^{\pi} P_{\max} G_{\left( \theta, \phi \right)} \, d\theta \, d\phi \\
& \approx \frac{P_{\max} \pi \theta_{3\text{dB}} \phi_{3\text{dB}}}{1.2 \ln10 \sqrt{1 - m^2 \theta_{3\text{dB}}^2 \phi_{3\text{dB}}^2 }},
\end{aligned}
\end{equation}
and we can obtain
\begin{equation}\label{Eq_3D_Pmax}
P_{\max} \approx P_t \frac{1.2 \ln10 \sqrt{1 - m^2 \theta_{3\text{dB}}^2 \phi_{3\text{dB}}^2 }}{\pi \theta_{3\text{dB}} \phi_{3\text{dB}}}.
\end{equation}
Substituting~\eqref{Eq_3D_Pmax} into~\eqref{Eq_3D_Pout_Case3_v}, we can calculate~\eqref{Eq_3D_Pout_Case3_v} as
\begin{equation}\label{Eq_3D_Pout_Case3_v_NEW}
\begin{aligned}
P_{\text{out}}^{3\text{D}, \infty} \left( \gamma_{th}, \mathbf{A} \right) = \Pr \left( \xi_1 v_1^2 + \xi_2 v_2^2  \geq h \left( \theta_{3\text{dB}}, \phi_{3\text{dB}}, m \right) \right)
\end{aligned}
\end{equation}
where
\begin{equation}\label{Eq_3D_h}
h \left( \theta_{3\text{dB}}, \phi_{3\text{dB}}, m \right) = \frac{d^2}{1.2} \lg \frac{1.2 \ln10 A_e P_t \sqrt{1 - m^2 \theta_{3\text{dB}}^2 \phi_{3\text{dB}}^2 }}{\pi \left( 4 \pi d \right)^2 \gamma_{th} \theta_{3\text{dB}} \phi_{3\text{dB}}}.
\end{equation}
To calculate $P_{\text{out}}^{3\text{D}, \infty} \left( \gamma_{th}, \mathbf{A} \right) $ in~\eqref{Eq_3D_Pout_Case3_v_NEW}, define $v_1 = r \cos \alpha, v_2 = r \sin \alpha$, and~\eqref{Eq_3D_Pout_Case3_v_NEW} can be calculated as
\begin{equation}\label{Eq_3D_Pout_Case3_r_alpha}
\begin{aligned}
&P_{\text{out}}^{3\text{D}, \infty} \left( \gamma_{th}, \mathbf{A} \right) \\
&= \! \iint \limits_{\substack{ \xi_1 v_1^2 + \xi_2 v_2^2 \\ \geq h \left( \theta_{3\text{dB}}, \phi_{3\text{dB}}, m \right) }} \! \frac{1}{2 \pi} \exp \left( - \frac{v_1^2 + v_2^2}{2} \right) dv_1 dv_2 \\
&= \int_0^{2 \pi} \int_{\sqrt{\frac{h \left( \theta_{3\text{dB}}, \phi_{3\text{dB}}, m \right)}{\xi_1 \cos^2 \alpha + \xi_2 \sin^2 \alpha}}}^{+\infty} \frac{r}{2 \pi} \exp \left( -\frac{r^2}{2} \right) dr d\alpha \\
&= \frac{1}{2 \pi} \int_0^{2 \pi} \exp \left( -\frac{h \left( \theta_{3\text{dB}}, \phi_{3\text{dB}}, m \right)}{2 \left( \xi_1 \cos^2 \alpha + \xi_2 \sin^2 \alpha \right)} \right) d\alpha.
\end{aligned}
\end{equation}
Since $P_{\text{out}}^{3\text{D}, \infty}$ in~\eqref{Eq_3D_Pout_Case3_r_alpha} involves multiple parameters, such as $\theta_{3\text{dB}}, \phi_{3\text{dB}}, m, \xi_1, \xi_2$, which complicates the analytical evaluation. To further simplify~\eqref{Eq_3D_Pout_Case3_r_alpha}, we need to calculate $h \left( \theta_{3\text{dB}}, \phi_{3\text{dB}}, m \right)$ in~\eqref{Eq_3D_h}. The determinants of $\mathbf{A}$ and $\mathbf{B}$ are given as
\begin{equation}\label{Eq_3D_det_A_B}
\left| \mathbf{A} \right| = \frac{1 - m^2 \theta_{3\text{dB}}^2 \phi_{3\text{dB}}^2}{\theta_{3\text{dB}}^2 \phi_{3\text{dB}}^2}, 
\left| \mathbf{B} \right| = \xi_1 \xi_2.
\end{equation}
According to~\eqref{Eq_3D_B}, we have
\begin{equation}\label{Eq_3D_A2B}
\left| \mathbf{B} \right| = \left| \mathbf{A} \right| \cdot \left| \mathbf{\Omega} \right| = \left| \mathbf{A} \right| \cdot | \mathbf{\mathbf{\Sigma}^{'2\text{D}}} |.
\end{equation}
Substituting~\eqref{Eq_3D_det_A_B} and~\eqref{Eq_3D_A2B} into~\eqref{Eq_3D_h}, we can compute $h \left( \theta_{3\text{dB}}, \phi_{3\text{dB}}, m \right) $ as
\begin{equation}\label{Eq_3D_h_xi1xi2}
\begin{aligned}
h \left( \xi_1, \xi_2 \right) &= \frac{d^2}{1.2} \lg \frac{1.2 \ln10   A_e P_t \left| \mathbf{A} \right|^{\frac{1}{2}}}{\pi \left( 4 \pi d \right)^2 \gamma_{th} } \\
&= \frac{d^2}{1.2} \lg \frac{1.2 \ln10   A_e P_t  \sqrt{\xi_1 \xi_2}}{\pi \left( 4 \pi d \right)^2 \gamma_{th} \left| \mathbf{\Sigma}^{'2\text{D}} \right|^{\frac{1}{2}}}.
\end{aligned}
\end{equation}
Through the above derivations, $h \left( \theta_{3\text{dB}}, \phi_{3\text{dB}}, m \right)$ in~\eqref{Eq_3D_h} can be equivalently reformulated as a function of the product $\xi_1 \xi_2$ in~\eqref{Eq_3D_h_xi1xi2}. Consequently, the outage probability in~\eqref{Eq_3D_Pout_Case3_r_alpha} depends only on $\xi_1, \xi_2$.

To calculate the optimal beam pattern, we first consider a special case where $\xi_1 \xi_2 = K$ with $K > 0$ being a constant. Hence, we need to solve the following optimization problem derived from~\eqref{Eq_3D_Pout_Case3_r_alpha} as
\begin{equation}\label{Eq_3D_Pout_Case3_xi1_equal_xi2}
\begin{aligned}
&P_{\text{out}}^{3\text{D}, \infty} \left( \gamma_{th}, \mathbf{A} \right) \\
&= \frac{1}{2 \pi} \int_0^{2 \pi} \exp \left( -\frac{h}{2 \left( \xi \cos^2 \alpha + \frac{K}{\xi} \sin^2 \alpha \right)} \right) d\alpha
\end{aligned}
\end{equation}
where we use $\xi$ to represent $\xi_1$ and $h = h \left( \xi_1, \xi_2 \right)$ is a constant in this case. 

It can be observed that the structure of~\eqref{Eq_3D_Pout_Case3_xi1_equal_xi2} is suitable for the application of Jensen’s inequality. Based on~\eqref{Eq_3D_Pout_Case3_xi1_equal_xi2}, we define
\begin{equation}\label{Eq_3D_g_xi}
g \left( \xi \right) = \frac{1}{2 \pi} \int_0^{2 \pi} \frac{d\alpha}{ \xi \cos^2 \alpha + \frac{K}{\xi} \sin^2 \alpha },
\end{equation}
and as demonstrated in Appendix~\ref{Appendix_g_xi_is_constant}, it remains constant with $g \left( \xi \right) = 1/\sqrt {K}$ for an arbitrary $\xi$. According to Jensen's inequality, we can compute~\eqref{Eq_3D_Pout_Case3_xi1_equal_xi2} as
\begin{equation}\label{Eq_3D_Jensen}
\begin{aligned}
P_{\text{out}}^{3\text{D}, \infty} \left( \gamma_{th}, \mathbf{A} \right)  &= \mathbb{E} \left[ \exp  \left( -\frac{h}{2 \left( \xi \cos^2 \alpha + \frac{K}{\xi} \sin^2 \alpha \right)} \right)  \right]  \\
& \geq \exp \left( -\frac{h}{2} \cdot \mathbb{E} \left[ \frac{1}{ \xi \cos^2 \alpha + \frac{K}{\xi} \sin^2 \alpha} \right] \right) \\
&= \exp \left( -\frac{h}{2} \cdot g \left( \xi \right) \right) =  \exp \left( -\frac{h}{2} \cdot \frac{1}{\sqrt{K}} \right)
\end{aligned}
\end{equation}
with equality if and only if $\xi \cos^2 \alpha + \frac{K}{\xi} \sin^2 \alpha$ is constant, which requires $\xi = \frac{K}{\xi}$ and furthermore, we have the optimal solution $\xi^\ast = \xi_1 = \xi_2 = \sqrt{K}$.

Equation~\eqref{Eq_3D_Jensen} shows that $P_{\text{out}}^{3\text{D}, \infty} \left( \gamma_{th}, \mathbf{A} \right) $ attains its minimum if and only if $\xi_1 = \xi_2$ under the constraint $\xi_1 \xi_2 = K$. In other words, any point that does not satisfy $\xi_1 = \xi_2$ cannot be a minimizer under this constraint, and thus cannot be a global minimizer. Consequently, the minimum of  $P_{\text{out}}^{3\text{D}, \infty} \left( \gamma_{th}, \mathbf{A} \right) $ is attained when $\xi_1 = \xi_2$. 
Substituting $\xi_1 = \xi_2$ into $\mathbf{M}$ in~\eqref{Eq_3D_B_Eigen}, we can calculate the optimal $\mathbf{B}^\ast$ as
\begin{equation}\label{Eq_3D_B_optimal}
\mathbf{B^\ast} = \mathbf{Q} 
\begin{bmatrix}
\xi^\ast & \\
 & \xi^\ast
\end{bmatrix}
\mathbf{Q}^T = \xi^\ast \mathbf{E}
\end{equation}
where $\mathbf{E}$ is identity matrix. Substituting~\eqref{Eq_3D_B_optimal} into~\eqref{Eq_3D_B}, we can compute the optimal $\mathbf{A}^\ast$ as
\begin{equation}\label{Eq_3D_A_optimal}
\mathbf{A}^\ast \!=\! \mathbf{G} \sqrt{\mathbf{\Omega}^{-1}} \mathbf{B}^\ast \sqrt{\mathbf{\Omega}^{-1}} \mathbf{G}^T \!=\! \xi^\ast \mathbf{G} \mathbf{\Omega}^{-1} \mathbf{G}^T \!=\! \xi^\ast \left( \mathbf{\Sigma}^{'2\text{D}} \right)^{-1}.
\end{equation}

Equation~\eqref{Eq_3D_A_optimal} indicates that the ratio between the major and minor axes of the beam and its rotation angle should be aligned with the positioning error in order to minimize the outage probability. This observation is insightful, as the beam achieves maximum coverage of the region with the highest probability of user distribution only when its shape is matched to the characteristics of the positioning error.

Subsequently, we need to compute the optimal value of $\xi^\ast$. Equation~\eqref{Eq_3D_Jensen} can be rewritten as the function of $\xi$ by substituting~\eqref{Eq_3D_h_xi1xi2} into~\eqref{Eq_3D_Jensen} as
\begin{equation}\label{Eq_3D_Pout_xi}
\begin{aligned}
P_{\text{out}}^{3\text{D}, \infty} \left( \xi \right)  &= \exp \left( -\frac{h \left( \xi \right)}{2 \xi} \right) \\
&= \exp \left( -\frac{d^2}{2.4 \xi} \lg \frac{1.2 \ln10   A_e P_t  \xi}{\pi \left( 4 \pi d \right)^2 \gamma_{th} \left| \mathbf{\Sigma}^{'2\text{D}} \right|^{\frac{1}{2}} }  \right).
\end{aligned}
\end{equation}
Taking the derivative of~\eqref{Eq_3D_Pout_xi} in terms of $\xi$ and making it zero, we can obtain the optimal solution of~\eqref{Eq_3D_Pout_xi} as
\begin{equation}\label{Eq_3D_xi_optimal}
\xi^\ast = \frac{e \pi \left( 4 \pi d \right)^2 \gamma_{th} \left| \mathbf{\Sigma}^{'2\text{D}} \right|^{\frac{1}{2}} }{1.2 \ln 10 A_e P_t}.
\end{equation}
According to~\eqref{Eq_3D_A},~\eqref{Eq_3D_Cov_Matrix_xz},~\eqref{Eq_3D_A_optimal} and~\eqref{Eq_3D_xi_optimal}, the optimal beam pattern in 3D scenarios can be calculated as

\begin{equation}\label{Eq_3D_optimal_beamwidth}
\begin{aligned}
\theta_{3\text{dB}}^\ast &= \sqrt{\frac{1.2 \ln 10 A_e P_t \left| \mathbf{\Sigma}^{'2\text{D}} \right|^{\frac{1}{2}} }{ e \pi \left( 4 \pi d \right)^2 \gamma_{th}  \mathbf{\Sigma}^{3\text{D}} \left( 3,3 \right)} }, \\
\phi_{3\text{dB}}^\ast &= \sqrt{\frac{1.2 \ln 10 A_e P_t \left| \mathbf{\Sigma}^{'2\text{D}} \right|^{\frac{1}{2}} }{ e \pi \left( 4 \pi d \right)^2 \gamma_{th}  \mathbf{\Sigma}^{3\text{D}} \left( 1,1 \right) } }, \\
m^\ast &= \frac{ e \pi \left( 4 \pi d \right)^2 \gamma_{th}  \mathbf{\Sigma}^{3\text{D}} \left( 1,3 \right)}{1.2 \ln 10 A_e P_t \left| \mathbf{\Sigma}^{'2\text{D}} \right|^{\frac{1}{2}} },
\end{aligned}
\end{equation}
and the lowest outage probability in~\eqref{Eq_3D_Pout_xi} is
\begin{equation}\label{Eq_3D_Pout_Optimal}
P_{\text{out}}^{3\text{D}, \ast}  = \exp \left( -\frac{A_e P_t}{2 e \pi \left( 4 \pi \right)^2 \gamma_{th} \left| \mathbf{\Sigma}^{'2\text{D}} \right|^{\frac{1}{2}} } \right).
\end{equation}
Notably, as $m$ is inherently the function of the beam rotation angle $\psi$, it is thus parameterized by $\psi$, which describes the smaller angle between the beam's major axis and the positive direction of the coordinate axes. Since $\mathbf{A}$ can be eigendecomposed as
\begin{equation}\label{Eq_3D_eigendecompose_A}
\mathbf{A} \! = \! \mathbf{U} \mathbf{Y} \mathbf{U}^T \! = \!
\begin{bmatrix}
\cos\psi & -\sin\psi \\
\sin\psi & \cos\psi
\end{bmatrix}
\begin{bmatrix}
\zeta_1 & 0 \\
0 & \zeta_2
\end{bmatrix}
\begin{bmatrix}
\cos\psi & \sin\psi \\
-\sin\psi & \cos\psi
\end{bmatrix} \! ,
\end{equation}
by combining~\eqref{Eq_3D_A},~\eqref{Eq_3D_optimal_beamwidth} and~\eqref{Eq_3D_eigendecompose_A}, we can compute $\psi^\ast$ as
\begin{equation}\label{Eq_3D_optimal_psi}
\psi^\ast = \frac{1}{2} \arctan \frac{2 \mathbf{\Sigma}^{3\text{D}} \left( 1,3 \right) }{\mathbf{\Sigma}^{3\text{D}} \left( 1,1 \right) - \mathbf{\Sigma}^{3\text{D}} \left( 3,3 \right) }.
\end{equation}
Based on the derived expression in~\eqref{Eq_3D_optimal_psi}, we proceed to analyze two special cases as follows:
\subsubsection{Case 1: $\mathbf{\Sigma}^{3\mathrm{D}} \left( 1,1 \right) = \mathbf{\Sigma}^{3\mathrm{D}} \left( 3,3 \right)$}
In this case, the positioning error exhibits equal variance along the two directions, and the contours of the error distribution form a circle. Hence, the beam pattern also becomes circular, and its rotation angle $\psi$ can be chosen arbitrarily. It is also worth noting that, $\theta_{3\text{dB}}$ and $\phi_{3\text{dB}}$ coincide in this case according to~\eqref{Eq_3D_optimal_beamwidth}.

\subsubsection{Case 2: $\psi = \pm \pi / 4$}
In~\eqref{Eq_3D_optimal_psi}, the rotation angle $\psi$ cannot reach $\pm \pi / 4$, although such a configuration may occur in practical implementations. Under this condition, the contours of the positioning error have equal intercepts along the $x$ and $y$-axes, which implies that $\theta_{3\text{dB}} = \phi_{3\text{dB}}$. This scenario is effectively encompassed by Case~1.

Based on~\eqref{Eq_2D_Optimal theta}, eq.~\eqref{Eq_3D_optimal_beamwidth} further extends the analysis by incorporating the statistical distribution of user positioning errors into the beam design framework. Specifically, it derives a closed-form expression for the optimal beamwidth that accounts for the angular dispersion introduced by positioning uncertainties. When the variance of the user distribution is large, the users are more likely to be dispersed over a wider angular range. In this case, a broader beam is required along the corresponding direction to maintain adequate coverage and ensure that a significant portion of the user probability mass falls within the main lobe of the beam pattern. Conversely, when the user distribution exhibits a smaller variance, indicating higher positioning accuracy, the beam can be more tightly focused. Narrower beamwidth enhances signal concentration, which can improve the received signal strength and reduce power leakage into undesired directions. In this way, the direction-adaptive strategy balances coverage and signal gain according to the spatial uncertainty of the users, which enables robust beamforming under imperfect user positioning information.

\section{Asymptotic Analysis of Approximation Error}\label{Asymptotic Error Analysis}
In this section, we investigate the approximation error between~\eqref{Eq_2D_Pout_Case3_New} and~\eqref{Eq_2D_Pout_Approximate} in 2D cases, and between~\eqref{Eq_3D_Pout_Case3_Origin} and~\eqref{Eq_3D_Pout_Case3_New} in 3D cases, while also analyze its asymptotic performance.
\subsection{Asymptotic Analysis of 2D Approximation Error}\label{Asymptotic Error Analysis_2D}
Define $\varepsilon^{2\text{D}}$ as the error between the exact outage probability in~\eqref{Eq_2D_Pout_Case3_New} and approximate outage probability in~\eqref{Eq_2D_Pout_Approximate}, which can be expressed according to the law of total probability as
\begin{equation}\label{Eq_2D_varepsilon}
	\begin{aligned}
		\varepsilon^{2\text{D}} &= P_{\text{out}}^{2\text{D}} \left( \gamma_{th}, \theta_{3 \text{dB}} \right) - P_{\text{out}}^{2\text{D}, \infty} \left( \gamma_{th}, \theta_{3 \text{dB}} \right) \\
		&= \Pr \left( \left| \hat{x}_u \right| \geq k \hat{y}_u \right) -  \Pr \left(\left|\hat{x}_u \right| \geq k d \right) \\
		&= \int_{-\infty}^{+\infty} \underbrace{\Pr \left( \left| \hat{x}_u \right| \geq k y \mid  \hat{y}_u = y \right)}_{p \left( y \right)} f_{\hat{y}_u} \left( y \right) dy \\
		& - \int_{-\infty}^{+\infty} \underbrace{\Pr \left( \left| \hat{x}_u \right| \geq k d \mid  \hat{y}_u = y \right)}_{q \left( y \right)} f_{\hat{y}_u} \left( y \right) dy \\
		&= \mathbb{E} \left[ p \left( \hat{y}_u \right) \right] - \mathbb{E} \left[ q \left( \hat{y}_u \right) \right]
	\end{aligned}
\end{equation}
where $f_{\hat{y}_u} \left( y \right)$ is the marginal PDF of $\hat{y}_u$ and $\mathbb{E} \left[ \cdot \right]$ represents the expectation of the function. To calculate $\mathbb{E} \left[ p \left( \hat{y}_u \right) \right]$ and $\mathbb{E} \left[ q \left( \hat{y}_u \right) \right]$, we perform Taylor series expansion of $p \left( \hat{y}_u \right)$ and $q \left( \hat{y}_u \right)$ around $\hat{y}_u = d$, given by
\begin{equation}\label{Eq_2D_Taylor_p}
	\begin{aligned}
		p \left( \hat{y}_u \right) &= p(d) + \left. \frac{d p \left( y \right)}{d y} \right|_{y= d} \cdot \left( \hat{y}_u - d \right) + \frac{1}{2} \cdot \left. \frac{d^2 p \left( y \right)}{d y^2} \right|_{y = d} \\
		&\quad \times \left( \hat{y}_u - d \right)^2 + o \left( \left( \hat{y}_u - d \right)^2 \right),  \hat{y}_u - d \rightarrow 0
	\end{aligned}
\end{equation}
and
\begin{equation}\label{Eq_2D_Taylor_q}
	\begin{aligned}
		q \left( \hat{y}_u \right) &= q(d) + \left. \frac{d q \left( y \right)}{d y} \right|_{y= d} \cdot \left( \hat{y}_u - d \right) + \frac{1}{2} \cdot \left. \frac{d^2 q \left( y \right)}{d y^2} \right|_{y = d} \\
		& \quad \times \left( \hat{y}_u - d \right)^2 + o \left( \left( \hat{y}_u - d \right)^2 \right),  \hat{y}_u - d \rightarrow 0.
	\end{aligned}
\end{equation}
Taking the expectation on both sides of~\eqref{Eq_2D_Taylor_p} and~\eqref{Eq_2D_Taylor_q}, $\mathbb{E} \left[ p \left( \hat{y}_u \right) \right]$ and $\mathbb{E} \left[ q \left( \hat{y}_u \right) \right]$ can be computed as
\begin{equation}\label{Eq_2D_Ep}
	\begin{aligned}
		\mathbb{E} \left[ p \left( \hat{y}_u \right) \right] &= p(d) +  \left. \frac{d p \left( y \right)}{d y} \right|_{y= d} \cdot \mathbb{E} \left[ \left( \hat{y}_u - d \right) \right] \\
		& + \frac{1}{2} \! \cdot \! \left. \frac{d^2 p \left( y \right)}{d y^2} \right|_{y = d} \!\! \cdot \! \mathbb{E} \left[ \! \left( \hat{y}_u - d \right)^2 \right] \! + \! o \left( \mathbb{E} \! \left[ \! \left( \hat{y}_u - d \right)^2 \right] \right)  \\
		&= p(d) + \frac{1}{2} \cdot \left. \frac{d^2 p \left( y \right)}{d y^2} \right|_{y = d} \cdot \sigma_y^2 + o \left( \sigma_y^2 \right), \sigma_y \rightarrow 0
	\end{aligned}
\end{equation}
and
\begin{equation}\label{Eq_2D_Eq}
	\begin{aligned}
		\mathbb{E} \left[ q \left( \hat{y}_u \right) \right] &= q(d) +  \left. \frac{d q \left( y \right)}{d y} \right|_{y= d} \cdot \mathbb{E} \left[ \left( \hat{y}_u - d \right) \right] \\
		& + \frac{1}{2} \! \cdot \! \left. \frac{d^2 q \left( y \right)}{d y^2} \right|_{y = d} \!\! \cdot \! \mathbb{E} \left[ \! \left( \hat{y}_u - d \right)^2 \right] \! + \! o \left( \mathbb{E} \! \left[ \! \left( \hat{y}_u - d \right)^2 \right] \right)  \\
		&= q(d) + \frac{1}{2} \cdot \left. \frac{d^2 q \left( y \right)}{d y^2} \right|_{y = d} \cdot \sigma_y^2 + o \left( \sigma_y^2 \right), \sigma_y \rightarrow 0.
\end{aligned}
\end{equation}
Since $p \left( d \right) = q \left( d \right)$, substituting~\eqref{Eq_2D_Ep} and~\eqref{Eq_2D_Eq} into~\eqref{Eq_2D_varepsilon}, $\varepsilon$ can be calculated as 
\begin{equation}\label{Eq_2D_varepsilon_p2d-q2d}
	\begin{aligned}
		\varepsilon^{2\text{D}} &= \mathbb{E} \left[ p \left( \hat{y}_u \right) \right] - \mathbb{E} \left[ q \left( \hat{y}_u \right) \right] \\
		&= \frac{\sigma_y^2}{2} \left( \left. \frac{d^2 p \left( y \right)}{d y^2} \right|_{y = d} - \left. \frac{d^2 q \left( y \right)}{d y^2} \right|_{y = d} \right) + o \left( \sigma_y^2 \right).
	\end{aligned}
\end{equation}
To derive the second-order derivative in~\eqref{Eq_2D_varepsilon_p2d-q2d}, the expressions of $p \left( y \right)$ and $q \left( y \right)$ must first be derived. Since $p \left( y \right)$ and $q \left( y \right)$ are defined as the conditional probability, we need to obtain the conditional PDF. Based on the formula for the conditional expectation and variance of bivariate Gaussian distribution, the conditional PDF of  $\hat{x}_u$ for a given $\hat{y}_u = y$ is
\begin{equation}
	\begin{aligned}
		f_{\hat{x}_u \mid \hat{y}_u} \left( x \mid y \right) = \frac{1}{\sqrt{2\pi} \sigma_{x \mid y}} \exp \left( -\frac{ \left( x - \mu_{x \mid y} \right)^2 }{ 2 \sigma_{x \mid y}^2} \right)
	\end{aligned}
\end{equation}
where $\mu_{x \mid y} = \frac{\rho \sigma_x}{\sigma_y} \left( y - d \right)$ and $\sigma_{x \mid y}^2 = \sigma_x^2 \left(1 - \rho^2 \right)$ is the conditional expectation and variance, and $\rho$ denotes the correlation coefficient between $\hat{x}_u$ and $\hat{y}_u$. Based on $\mu_{x \mid y}$ and $\sigma_{x \mid y}^2$, we can calculate $p \left( y \right)$ and $q \left( y \right)$ as
\begin{equation}\label{Eq_2D_p}
	\begin{aligned}
		p \left( y \right) &= \Pr \left( \left| \hat{x}_u \right| \geq k y \mid \hat{y}_u = y \right) \\
		&= \Pr \left( \hat{x}_u \geq k y \mid \hat{y}_u = y \right) + \Pr \left(  \hat{x}_u  < - k y \mid \hat{y}_u = y \right) \\
		&= Q \left( \frac{k  y - \mu_{x \mid y}}{\sigma_{x \mid y}} \right) + 1 - Q \left( \frac{- k  y - \mu_{x \mid y}}{\sigma_{x \mid y}} \right)  \\
		&= Q \left( \frac{k  y - \frac{\rho \sigma_x}{\sigma_y} \left( y - d \right)}{\sigma_x \sqrt{1 - \rho^2}} \right) + Q \left( \frac{k  y + \frac{\rho \sigma_x}{\sigma_y} \left( y - d \right)}{\sigma_x \sqrt{1 - \rho^2}} \right)
	\end{aligned}
\end{equation}
and
\begin{equation}\label{Eq_2D_q}
	\begin{aligned}
		q \left( y \right) &= \Pr \left( \left| \hat{x}_u \right| \geq k d \mid \hat{y}_u = y \right) \\
		&= \Pr \left( \hat{x}_u \geq k d \mid \hat{y}_u = y \right) + \Pr \left(  \hat{x}_u  < - k d \mid \hat{y}_u = y \right) \\
		&= Q \left( \frac{k  d - \mu_{x \mid y}}{\sigma_{x \mid y}} \right) + 1 - Q \left( \frac{- k  d - \mu_{x \mid y}}{\sigma_{x \mid y}} \right)  \\
		&= Q \left( \frac{k  d - \frac{\rho \sigma_x}{\sigma_y} \left( y - d \right)}{\sigma_x \sqrt{1 - \rho^2}} \right) + Q \left( \frac{k  d + \frac{\rho \sigma_x}{\sigma_y} \left( y - d \right)}{\sigma_x \sqrt{1 - \rho^2}} \right).
	\end{aligned}
\end{equation}
Taking the second-order derivative of  $p \left( \hat{y}_u \right)$ and $q \left( \hat{y}_u \right)$ in~\eqref{Eq_2D_p} and~\eqref{Eq_2D_q} and substituting them into~\eqref{Eq_2D_varepsilon_p2d-q2d}, $\varepsilon$ can be computed as
\begin{equation}\label{Eq_2D_varepsilon_final}
	\varepsilon^{2\text{D}} \approx \frac{k^3 d \sigma_y^2 }{\sqrt{2 \pi} \sigma_x^3 \left( 1 - \rho^2 \right)^{3/2}} \exp{ \left( -\frac{k^2 d^2}{2 \sigma_x^2 \left( 1 - \rho^2 \right)} \right) }.
\end{equation}
The asymptotic behavior of $\varepsilon^{2\text{D}}$ can be analyzed in two cases.
\subsubsection{Case 1: $d \rightarrow +\infty$}
In this case, the exponential term in~\eqref{Eq_2D_varepsilon_final} satisfies
\begin{equation}
\lim_{d \rightarrow +\infty} \exp{ \left( -\frac{k^2 d^2}{2 \sigma_x^2 \left( 1 - \rho^2 \right)} \right) } = 0.
\end{equation}
Since the exponential decays much faster than the linear term in~\eqref{Eq_2D_varepsilon_final}, we have
\begin{equation}
\lim_{d \rightarrow +\infty} \varepsilon^{2\text{D}} = 0.
\end{equation}
\subsubsection{Case 2: $\rm{tr} \left( \mathbf{\Sigma}^{2\mathrm{D}} \right) \rightarrow 0$}
This case is equivalent to $\sigma_x \rightarrow 0$ and $\sigma_y \rightarrow 0$, so we have
\begin{equation}
\lim_{\rm{tr} \left( \mathbf{\Sigma}^{2\text{D}} \right) \rightarrow 0} \exp{ \left( -\frac{k^2 d^2}{2 \sigma_x^2 \left( 1 - \rho^2 \right)} \right) } = 0.
\end{equation}
For the same reason in case 1, we can conclude that
\begin{equation}
\lim_{\rm{tr} \left(\mathbf{\Sigma}^{2\text{D}} \right) \rightarrow 0} \varepsilon^{2\text{D}} = 0. 
\end{equation}

Case 1 corresponds to the scenario where the transmitter and the receiver are sufficiently far away, while Case 2 corresponds to the condition where the positioning system is accurate.

\subsection{Asymptotic Analysis of 3D Approximation Error}\label{Asymptotic Error Analysis_3D}
Define the error between the exact outage probability in~\eqref{Eq_3D_Pout_Case3_Origin} and approximate outage probability in~\eqref{Eq_3D_Pout_Case3_New} as
\begin{equation}\label{Eq_3D_varepsilon}
	\begin{aligned}
		\varepsilon^{3\text{D}} &= P_{\text{out}}^{3\text{D}} \left( \gamma_{th}, \mathbf{A} \right) - P_{\text{out}}^{3\text{D}, \infty} \left( \gamma_{th}, \mathbf{A} \right)  \\ 
		&= \Pr \left( \left[ \theta, \phi \right] \mathbf{A} \left[ \theta, \phi \right]^T \geq  \frac{1}{1.2} \lg \frac{P_{\max} A_e}{\left( 4 \pi d \right)^2 \gamma_{th}} \right) \\
		&- \Pr \left( \left[ \hat{x}_u, \hat{z}_u \right] \mathbf{A} \left[ \hat{x}_u, \hat{z}_u \right]^T \geq \frac{d^2}{1.2} \lg \frac{P_{\max} A_e}{\left( 4 \pi d \right)^2 \gamma_{th}} \right) \\
		&= \iiint\limits_{\substack{\left[ \theta, \phi \right] \mathbf{A} \left[ \theta, \phi \right]^T \\ \geq \frac{1}{1.2} \lg \frac{P_{\max} A_e}{\left( 4 \pi d \right)^2 \gamma_{th}}}} f_{\hat{x}_u,\hat{y}_u,\hat{z}_u} \left( x,y,z \right) dx dy dz \\
		&- \iiint\limits_{\substack{ \frac{1}{d^2} \left[ x, z \right] \mathbf{A} \left[ x,z \right]^T \\ \geq \frac{1}{1.2} \lg \frac{P_{\max} A_e}{\left( 4 \pi d \right)^2 \gamma_{th}}}} f_{\hat{x}_u,\hat{y}_u,\hat{z}_u} \left( x,y,z \right) dx dy dz.
	\end{aligned}
\end{equation}
The asymptotic performance of~\eqref{Eq_3D_varepsilon} can be analyzed in two cases.
\subsubsection{Case 1: $d \rightarrow +\infty$}
In this case, the right-hand sides of the inequalities, i.e., $\frac{1}{1.2} \lg \frac{P_{\max} A_e}{\left( 4 \pi d \right)^2 \gamma_{th}}$ in~\eqref{Eq_3D_varepsilon} tend to $-\infty$, while the left-hand sides, i.e., $\left[ \theta, \phi \right] \mathbf{A} \left[ \theta, \phi \right]^T$ and $\frac{1}{d^2} \left[ x, z \right] \mathbf{A} \left[ x,z \right]^T$ are positive. Consequently, the two inequalities always hold and thus the two integral regions extend to the entire 3D Euclidean space $\mathbb{R}^3$, which leads to
\begin{equation}
	\begin{aligned}
		\lim_{d \rightarrow +\infty} \varepsilon^{3\text{D}} &= \iiint_{\mathbb{R}^3} f_{\hat{x}_u,\hat{y}_u,\hat{z}_u} \left( x,y,z \right) dx dy dz \\
		&- \iiint_{\mathbb{R}^3} f_{\hat{x}_u,\hat{y}_u,\hat{z}_u} \left( x,y,z \right) dx dy dz \\
		&= 0.
	\end{aligned}
\end{equation}
\subsubsection{Case 2: $\rm{tr} \left( \mathbf{\Sigma}^{3\mathrm{D}} \right) \rightarrow 0$}
When $\rm{tr} \left( \mathbf{\Sigma}^{3\mathrm{D}} \right) \rightarrow 0$, the PDF function $f_{\hat{x}_u,\hat{y}_u,\hat{z}_u} \left( x,y,z \right)$ behaves like the Dirac delta distribution $\delta \left( x, y - d, z \right)$. For any region $\mathcal{D} \subseteq \mathbb{R}^3$ whose boundary $\partial \mathcal{D}$ does not contain the mean point $\left( 0,d,0 \right)$, it holds that~\cite{patrick1999convergence}
\begin{equation}
	\lim_{\mathbf{\Sigma}^{3\mathrm{D}}\to 0}\iiint_{\mathcal{D}} f_{\hat{x}_u,\hat{y}_u,\hat{z}_u}(x,y,z) dxdydz =
	\begin{cases}
		1, \!\!\quad (0,d,0) \in \mathcal{D},\\
		0, \!\!\quad (0,d,0) \notin \mathcal{D}.
	\end{cases}
\end{equation}
Since $\theta,\phi$ are functions of $\hat{x}_u,\hat{y}_u,\hat{z}_u$ according to~\eqref{Eq_3D_theta} and~\eqref{Eq_3D_phi}, evaluating them at the mean point $(0,d,0)$ gives $\left[ \theta, \phi \right] \mathbf{A} \left[ \theta, \phi \right]^T = \frac{1}{d^2} \left[ x, z \right] \mathbf{A} \left[ x,z \right]^T = 0$. Hence, if $(0,d,0)$ is contained by one of the integral region in~\eqref{Eq_3D_varepsilon}, it must be contained by the other. In this way, the two integrals in~\eqref{Eq_3D_varepsilon} simultaneously take the value 0 or 1, and thus
\begin{equation}
	\begin{aligned}
		\lim_{\rm{tr} \left( \mathbf{\Sigma}^{3\mathrm{D}} \right) \rightarrow 0} \varepsilon^{3\text{D}}&= \iiint\limits_{\substack{\left[ \theta, \phi \right] \mathbf{A} \left[ \theta, \phi \right]^T \\ \geq \frac{1}{1.2} \lg \frac{P_{\max} A_e}{\left( 4 \pi d \right)^2 \gamma_{th}}}} f_{\hat{x}_u,\hat{y}_u,\hat{z}_u} \left( x,y,z \right) dx dy dz \\
		&- \iiint\limits_{\substack{ \frac{1}{d^2} \left[ x, z \right] \mathbf{A} \left[ x,z \right]^T \\ \geq \frac{1}{1.2} \lg \frac{P_{\max} A_e}{\left( 4 \pi d \right)^2 \gamma_{th}}}} f_{\hat{x}_u,\hat{y}_u,\hat{z}_u} \left( x,y,z \right) dx dy dz.\\
		&= 0.
	\end{aligned}
\end{equation}

Similar to the 2D scenario, Case 1 and Case 2 correspond to the situations where the link distance approaches infinity and the positioning accuracy becomes extremely high, respectively.

\section{Simulation Results}\label{Simulation Results}
In this section, we present simulation results for the outage probability of both 2D and 3D beamforming systems based on positioning information under various parameter settings.

\begin{figure}
	\centering
	\vspace{-8pt}
	\includegraphics[width=0.45\textwidth]{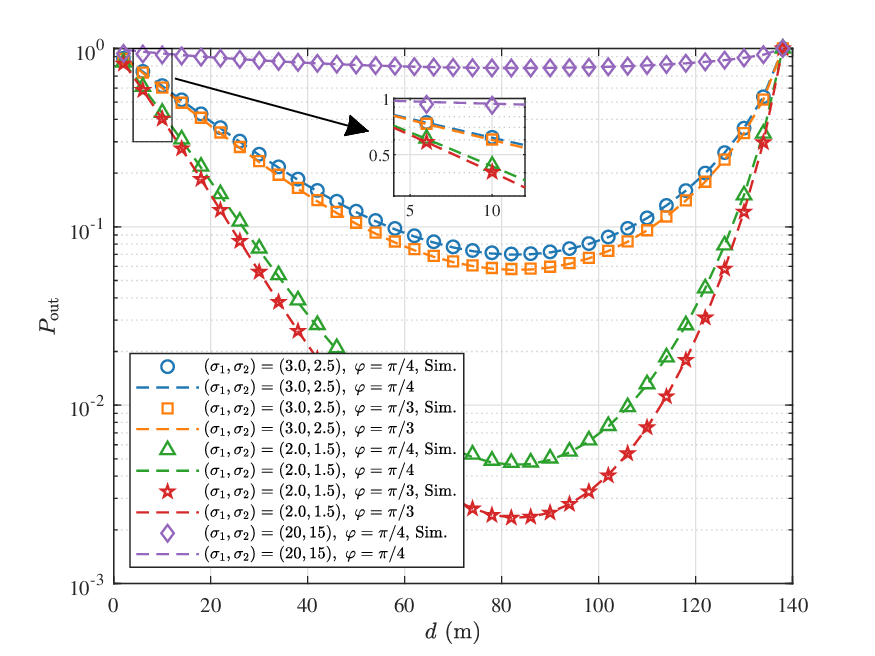}\\
	\caption{Outage probability as a function of the distance $d$ for 2D beamforming systems. $P_t = 25$~dB; $A_e = 1$~cm$^2$; $a_m = 10^{-4}$; $\gamma_{th} = 10^{-7}$~W; $\theta_{3 \text{dB}} = 0.1$~rad.}\label{Figure_2D_d}
\end{figure}

Figure~\ref{Figure_2D_d} depicts the outage probability versus distance $d$ for 2D beamforming systems under varying positioning error. It can be observed that the theoretical and simulated outage probabilities exhibit a close match, which validates the accuracy of the analytical derivations. The results indicate that the outage probability remains low when the distance is either small or large, while it peaks at intermediate distance. This behavior arises because insufficient received signal power due to severe path loss occurs at larger distance, whereas inadequate coverage of the angular spread by the fixed beamwidth at shorter distance leads to increased misalignment. Furthermore, the outage probability tends to decrease with less positioning error, and it is also affected by the directional angle. It is worth noting that the approximation in~\eqref{Eq_2D_Pout_Approximate} is expected to lose accuracy when $d$ is small or $\rm{tr} \left( \mathbf{\Sigma}^{2\mathrm{D}} \right)$ becomes large, as discussed in Section~\ref{Asymptotic Error Analysis_2D}. However, the purple curve with extremely large positioning error and the zoomed-in view in Fig.~\ref{Figure_2D_d} indicate that the theoretical and simulated curves still closely overlap even under small $d$ and large positioning errors, which is consistent with the 2D asymptotic analysis in Section~\ref{Asymptotic Error Analysis_2D}.

\begin{figure}
	\centering
	\vspace{-8pt}
	\includegraphics[width=0.45\textwidth]{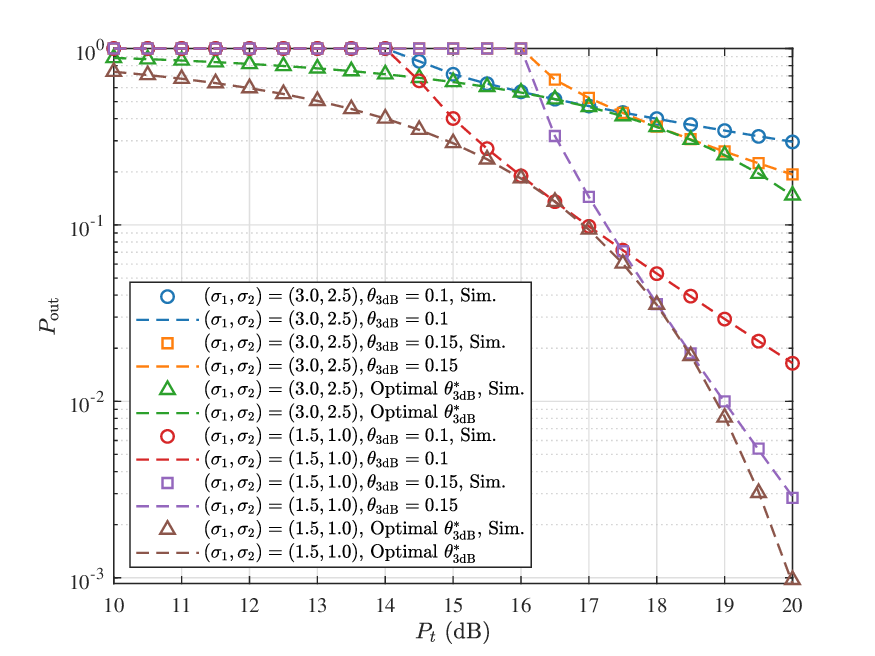}\\
	\caption{Outage probability as a function of the transmit power $P_t$ for 2D beamforming systems. $d = 40$~m; $A_e = 1$~cm$^2$; $a_m = 10^{-4}$; $\gamma_{th} = 10^{-7}$~W; $\varphi = \pi / 3$~rad.}\label{Figure_2D_Pt}
\end{figure}

Figure~\ref{Figure_2D_Pt} illustrates the impact of beamwidth and positioning error on the outage probability under different transmit power $P_t$. It can be observed that the outage probability consistently decreases as the transmit power increases. For a given positioning error, the optimal beamwidth achieves the minimum outage probability over the entire range of transmit power, and consistently outperforms non-optimal beamwidth configurations. For example, under the positioning error characterized by $\sigma_1 = 1.5, \sigma_2 = 1.0$, the outage probability curve corresponding to the optimal beamwidth becomes tangent to those of fixed beamwidth of 0.1~rad and~0.15 rad at certain transmit power levels, indicating that they achieve equal outage probability at the tangency points. However, outside these specific points, the optimal beamwidth consistently achieves a lower outage probability. This is because the fixed beamwidth coincides with the optimal value of~\eqref{Eq_2D_Optimal theta} at the tangency points, and thus achieving the minimum outage probability.

\begin{figure}
	\centering
	\vspace{-8pt}
	\includegraphics[width=0.45\textwidth]{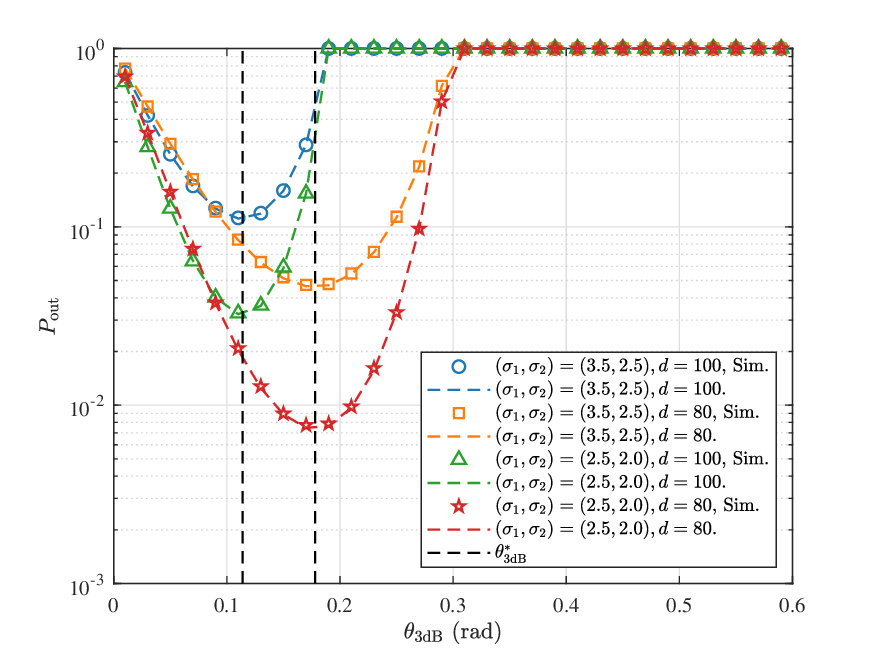}\\
	\caption{Outage probability as a function of the 3-dB beamwidth $\theta_{3 \text{dB}}$ for 2D beamforming systems. $P_t = 25$~dB; $A_e = 1$~cm$^2$; $a_m = 10^{-4}$; $\gamma_{th} = 10^{-7}$~W; $\varphi = \pi / 4$~rad.}\label{Figure_2D_theta}
\end{figure}

Figure~\ref{Figure_2D_theta} exhibits the outage probability as a function of $\theta_{3\text{dB}}$ under various distance and positioning error, assuming a fixed directional angle. The vertical lines in the figure indicate the optimal $\theta_{3\text{dB}}^\ast$ computed from~\eqref{Eq_2D_Optimal theta}, which precisely coincide with the minimum of the outage probability curves, and thus confirming the correctness of the theoretical result. Meanwhile, it reveals an important conclusion that, under a fixed beam direction, the outage probability is determined exclusively by the transmission distance, regardless of the positioning error, which not only aligns with the analytical result in~\eqref{Eq_2D_Optimal theta}, but also offers valuable guidance for optimal beamwidth design.

\begin{figure}
	\centering
	\vspace{-8pt}
	\includegraphics[width=0.45\textwidth]{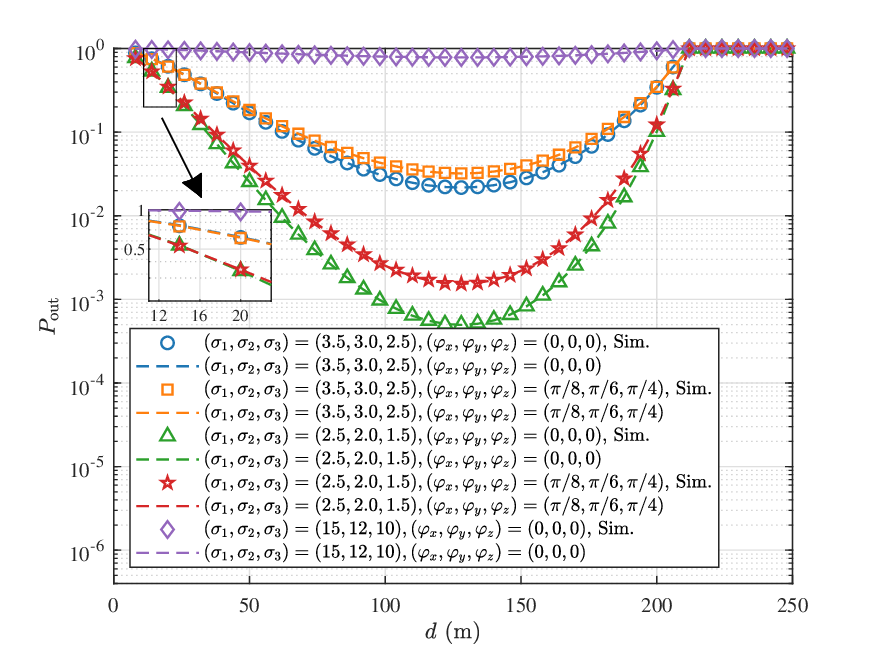}\\
	\caption{Outage probability as a function of the distance $d$ for 3D beamforming systems. $P_t = 20$~dB; $A_e = 1$~cm$^2$; $a_m = 10^{-4}$; $\gamma_{th} = 10^{-7}$~W; $\theta_{3 \text{dB}} = 0.1$~rad; $\phi_{3 \text{dB}} = 0.1$~rad; $\psi = 0.5$~rad.}\label{Figure_3D_d}
\end{figure}

Subsequently, we investigate the relationship between outage probability and distance $d$ under various positioning error patterns in 3D scenario and show them in Fig.~\ref{Figure_3D_d}. Consistent with 2D scenario, the close agreement observed between theoretical and simulated outage probability in the 3D case validates the accuracy of the analytical expressions. The outage probability versus distance exhibits a characteristic valley-shaped trend characterized by an initial decrease to a minimum followed by a monotonic increase at larger distance. This non-monotonic behavior arises from identical physical mechanisms to 2D case, wherein severe path loss dominates signal degradation at extended ranges while beam misalignment due to inadequate angular coverage of the fixed beamwidth relative to the positioning error becomes the primary impairment at proximal distance. Furthermore, both positioning error and directional angle significantly influence outage probability, confirming their criticality in system optimization. Similar to the 2D case, even when $d$ is small and $\rm{tr} \left( \mathbf{\Sigma}^{3\mathrm{D}} \right)$ is extremely large, both the purple curve with large positioning error and the zoomed-in view show that the theoretical and simulated curves remain in close agreement, which is consistent with the 3D asymptotic analysis in Section~\ref{Asymptotic Error Analysis_3D}.

\begin{figure}
	\centering
	\vspace{-8pt}
	\includegraphics[width=0.45\textwidth]{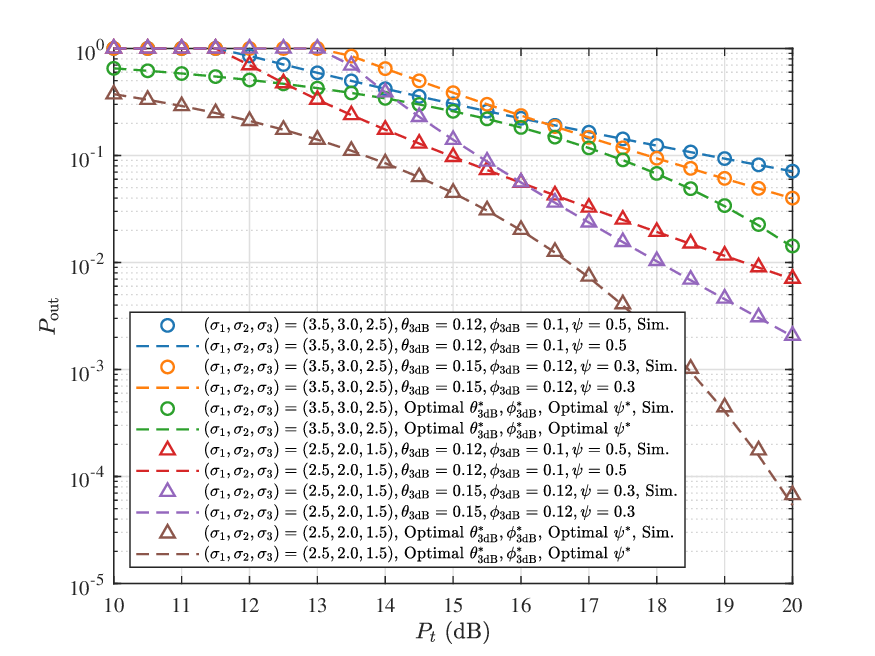}\\
	\caption{Outage probability as a function of the transmit power $P_t$ for 3D beamforming systems. $d = 80$~m; $A_e = 1$~cm$^2$; $a_m = 10^{-4}$; $\gamma_{th} = 10^{-7}$~W; $\left( \varphi_x, \varphi_y, \varphi_z \right) = \left( \pi/3, \pi/6, \pi/4 \right)$.}\label{Figure_3D_Pt}
\end{figure}

Figure~\ref{Figure_3D_Pt} shows the outage probability as a function of transmit power $P_t$ under various positioning error and beam rotation angle. While the overall trend resembles that observed in 2D case, the outage probability curves corresponding to non-optimal beamwidth consistently lie above that of the theoretically optimal beamwidth. This behavior arises because, in 3D setting, the optimal outage performance depends jointly on both the $\theta_{3 \text{dB}}$, $\phi_{3 \text{dB}}$ and $\psi$. When one of $\theta_{3 \text{dB}}$, $\phi_{3 \text{dB}}$ or $\psi$ is fixed, the jointly optimal condition cannot be fully satisfied across the entire range of transmit powers, and thus the curves do not exhibit tangency.

\begin{figure}
	\centering
	\vspace{-8pt}
	\includegraphics[width=0.45\textwidth]{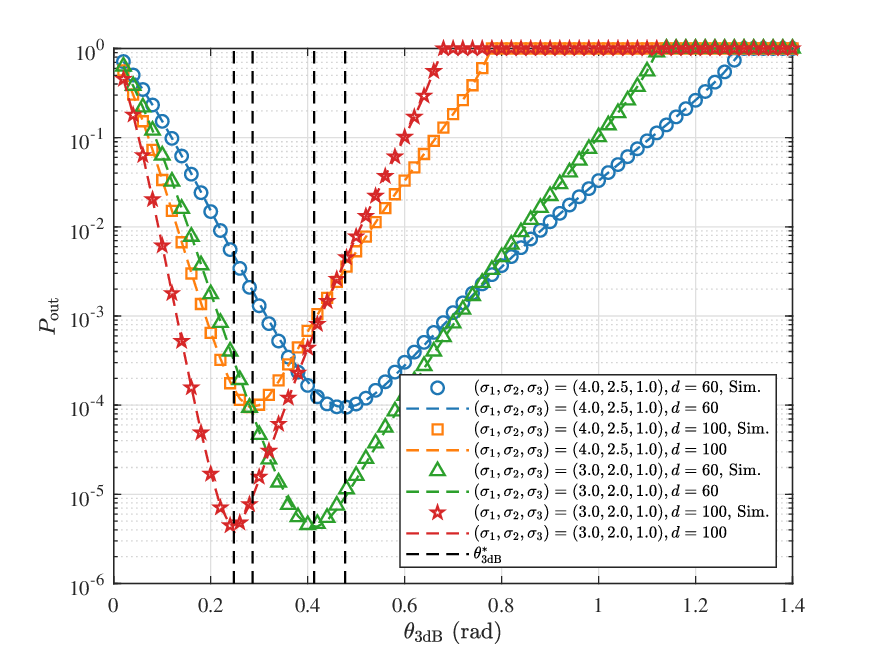}\\
	\caption{Outage probability as a function of the 3-dB beamwidth $\theta_{3 \text{dB}}$ for 3D beamforming systems. $P_t = 20$~dB; $A_e = 1$~cm$^2$; $a_m = 10^{-4}$; $\gamma_{th} = 10^{-7}$~W; Optimal $\phi_{3 \text{dB}}^\ast$; Optimal $\psi^\ast$; $\left( \varphi_x, \varphi_y, \varphi_z \right) = \left( 0, 0, 0 \right)$.}\label{3D_theta_NO1}
\end{figure}

Figure~\ref{3D_theta_NO1} characterizes the relationship between 3-dB beamwidth $\theta_{3 \text{dB}}$ and outage probability under fixed directional angle for varying positioning error and distance in 3D scenario. Although the optimal beamwidth in 2D scenario is independent of the positioning error, this is no longer the case in 3D setting, where the optimal beamwidth is significantly influenced by the spatial characteristics of the positioning error. The four curves correspond to distinct parameter sets where each configuration differs in either the positioning error or the distance, resulting in unique optimal beamwidth for all cases. For instance, both the green and blue curves correspond to the same value of $ \mathbf{\Sigma}^{3\text{D}} \left( 3,3 \right)$, meaning the optimal beamwidth primarily depends on the determinant of $\mathbf{\Sigma}^{'2\text{D}}$. Since the green curve has a smaller determinant than the blue one, it results in a smaller optimal beamwidth. The same dependency is also observed in other two curves, reinforcing the conclusion that the beamwidth is influenced by the positioning error's spatial distribution. This observation is in full agreement with the analytical expression provided in~\eqref{Eq_3D_optimal_beamwidth}.

\begin{figure}
	\centering
	\vspace{-8pt}
	\includegraphics[width=0.45\textwidth]{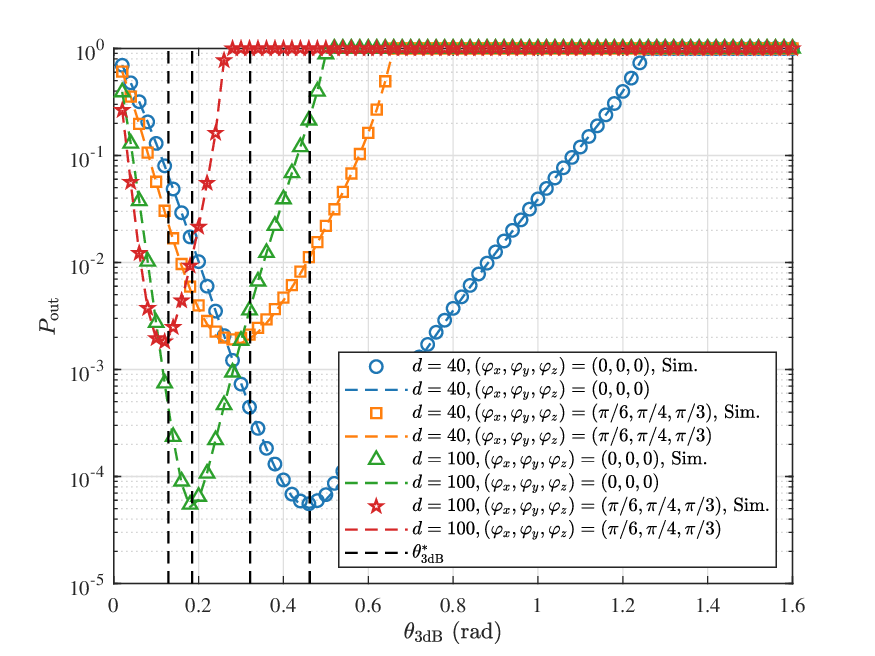}\\
	\caption{Outage probability as a function of the 3-dB beamwidth $\theta_{3 \text{dB}}$ for 3D beamforming systems. $P_t = 20$~dB; $A_e = 1$~cm$^2$; $a_m = 10^{-4}$; $\gamma_{th} = 10^{-7}$~W; Optimal $\phi_{3 \text{dB}}^\ast$; Optimal $\psi^\ast$; $\left( \sigma_1, \sigma_2, \sigma_3 \right) = \left( 2.5, 2.0, 1.5 \right)$.}\label{3D_theta_NO2}
\end{figure}

It can be observed from~\eqref{Eq_3D_optimal_beamwidth} that the optimal beamwidth depends on the positioning error and directional angle. Hence, Fig.~\ref{3D_theta_NO2} compares the outage probability and 3-dB beamwidth $\theta_{3 \text{dB}}$ under different directional angles while keeping the positioning error fixed. It can be observed that, with all other parameters held constant, variations in the rotation angle influence both the outage probability and the optimal beamwidth, which is also consistent with the result in~\eqref{Eq_3D_optimal_beamwidth}. Notably, as $\phi_{3 \text{dB}}$ exhibits the same trend as $\theta_{3 \text{dB}}$, its corresponding simulation results are not presented here to avoid redundancy.

\begin{figure}
	\centering
	\includegraphics[width=0.45\textwidth]{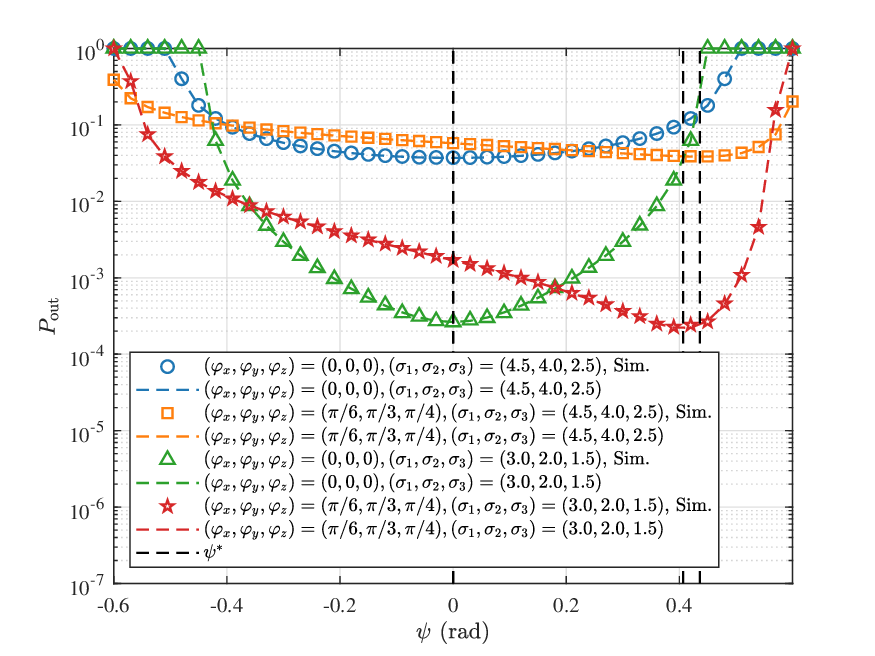}\\
	\caption{Outage probability as a function of the beam rotation angle $\psi$ for 3D beamforming systems. $d = 120$~m; $P_t = 20$~dB; $A_e = 1$~cm$^2$; $a_m = 10^{-4}$; $\gamma_{th} = 10^{-7}$~W; Optimal $\theta_{3 \text{dB}}^\ast$; Optimal $\phi_{3 \text{dB}}^\ast$.}\label{Figure_3D_m}
\end{figure}

Figure~\ref{Figure_3D_m} exhibits the outage probability as a function of beam rotation angle $\psi$ under various positioning error and directional angle. The results in the figure corroborate the conclusion in~\eqref{Eq_3D_optimal_psi}, which indicates that the principal axes of the beam must be aligned with those of the positioning error in order to cover the region with the highest user distribution probability. In addition, the optimal beam rotation angle depends on the magnitude of the positioning error. For instance, when the directional angle of the positioning error is zero, the corresponding error contours form an ellipse whose principal axes coincide with the coordinate axes, and no beam rotation is required, as illustrated by the blue and green curves in the figure. When the directional angle of the positioning error varies, the beam rotation angle should be adjusted accordingly to maintain the optimal alignment to preserve the optimal link quality.

\section{Conclusion}\label{Conclusion}
We investigated the positioning-assisted beamforming systems and analyze their performance for both 2D and 3D scenarios, where closed-form asymptotic expressions of the outage probability were derived based on the positioning error distribution, link distance, transmission power and beamwidth. The converging speed of the new asymptotic outage probability expressions was also quantified. Furthermore, we also developed the optimal 2D and 3D beam pattern expressions, laying a theoretical foundation for the design of the real beamforming systems incorporating positioning information. 

{\appendices
\section{}\label{Appendix_g_xi_is_constant}
We need to prove that~\eqref{Eq_3D_g_xi} is identically equal to $1/{\sqrt{K}}$ for an arbitrary $\xi$. Since the integral region is periodic with period $\pi$ and exhibits symmetry, we have
\begin{equation}\label{Eq_3D_g_xi_quarter}
g \left( \xi \right) = \frac{2}{\pi} \int_0^{\pi/2} \frac{d\alpha}{ \xi \cos^2 \alpha + \frac{K}{\xi} \sin^2 \alpha }.
\end{equation}
Define $\omega = \tan \alpha$, and~\eqref{Eq_3D_g_xi_quarter} can be computed as
\begin{equation}\label{Eq_3D_g_xi_omega}
\begin{aligned}
g \left( \xi \right) &= \frac{2}{\pi} \int_0^{+\infty} \frac{\omega^2 + 1}{\xi + \frac{K}{\xi} \omega^2} \cdot \frac{d\omega}{\omega^2 + 1}  \\
&= \frac{2 \xi}{\pi \sqrt{K}} \int_0^{+\infty} \frac{d \left( \sqrt{K}\omega \right)}{\xi^2 +  \left( \sqrt{K}\omega \right)^2} \\
&= \frac{2 \xi}{\pi \sqrt{K}} \cdot \frac{1}{\xi} \cdot \arctan \frac{\sqrt{K} \omega}{\xi} \bigg|^{+\infty}_0 \\
&= \frac{1}{\sqrt{K}}.
\end{aligned}
\end{equation}
Hence, the original claim is proved.


\bibliographystyle{IEEEtran}
\bibliography{REF}

\end{document}